\documentclass[journal=jacsat,manuscript=article]{achemso}

\usepackage[version=3]{mhchem} 
\usepackage{subfiles}
\usepackage{xr}
\makeatletter
\newcommand*{\addFileDependency}[1]{
  \typeout{(#1)}
  \@addtofilelist{#1}
  \IfFileExists{#1}{}{\typeout{No file #1.}}
}
\makeatother

\newcommand*{\myexternaldocument}[1]{
    \externaldocument{#1}
    \addFileDependency{#1.tex}
    \addFileDependency{#1.aux}
}

\myexternaldocument{Supplementary}

\author{Hao Zhang}
\affiliation[ECE]
{Department of Electrical and Computer Engineering, University of Victoria, Victoria, Canada}
\alsoaffiliation[CAMTEC]
{Centre for Advanced Materials $\&$ Related Technologies (CAMTEC), University of Victoria, Victoria, Canada}

\author{Parinaz Moazzezi}
\affiliation[ECE]
{Department of Electrical and Computer Engineering, University of Victoria, Victoria, Canada}
\alsoaffiliation[CAMTEC]
{Centre for Advanced Materials $\&$ Related Technologies (CAMTEC), University of Victoria, Victoria, Canada}

\author{Juanjuan Ren}
\affiliation{Department of Physics, Engineering Physics and Astronomy, Queen's University, Kingston, Canada}

\author{Brett Henderson}
\affiliation[chem]
{Department of Chemistry, University of Victoria, Victoria, Canada}
\alsoaffiliation[CAMTEC]
{Centre for Advanced Materials $\&$ Related Technologies (CAMTEC), University of Victoria, Victoria, Canada}
\alsoaffiliation[company]
{Quantum Algorithms Institute, Surrey, Canada}

\author{Cristina Cordoba}
\affiliation[phys]
{Department of Physics and Astronomy, University of Victoria, Victoria, Canada}
\alsoaffiliation[CAMTEC]
{Centre for Advanced Materials $\&$ Related Technologies (CAMTEC), University of Victoria, Victoria, Canada}

\author{Vishal Yeddu}
\affiliation[chem]
{Department of Chemistry, University of Victoria, Victoria, Canada}
\alsoaffiliation[CAMTEC]
{Centre for Advanced Materials $\&$ Related Technologies (CAMTEC), University of Victoria, Victoria, Canada}

\author{Arthur Blackburn}
\affiliation[phys]
{Department of Physics and Astronomy, University of Victoria, Victoria, Canada}
\alsoaffiliation[CAMTEC]
{Centre for Advanced Materials $\&$ Related Technologies (CAMTEC), University of Victoria, Victoria, Canada}

\author{Makhsud I. Saidaminov}
\affiliation[ECE]
{Department of Electrical and Computer Engineering, University of Victoria, Victoria, Canada}
\alsoaffiliation[CAMTEC]
{Centre for Advanced Materials $\&$ Related Technologies (CAMTEC), University of Victoria, Victoria, Canada}
\alsoaffiliation[chem]
{Department of Chemistry, University of Victoria, Victoria, Canada}

\author{Irina Paci}
\affiliation[chem]
{Department of Chemistry, University of Victoria, Victoria, Canada}
\alsoaffiliation[CAMTEC]
{Centre for Advanced Materials $\&$ Related Technologies (CAMTEC), University of Victoria, Victoria, Canada}

\author{Stephen Hughes}
\affiliation{Department of Physics, Engineering Physics and Astronomy, Queen's University, Kingston, Canada}

\author{Reuven Gordon}
\affiliation[ECE]
{Department of Electrical and Computer Engineering, University of Victoria, Victoria, Canada}
\alsoaffiliation[CAMTEC]
{Centre for Advanced Materials $\&$ Related Technologies (CAMTEC), University of Victoria, Victoria, Canada}
\email{rgordon@uvic.ca}
\phone{+1 (250) 472-5179}
\fax{+1 (250) 721-6052}

\title[Coupling Perovskite Quantum Dot Pairs in Solution using Nanoplasmonic Assembly]
  {
Coupling Perovskite Quantum Dot Pairs in Solution using Nanoplasmonic Assembly}

\abbreviations{IR,NMR,UV}

\begin{document}







\begin{abstract}
  Perovskite quantum dots (PQDs) provide a robust solution-based approach to efficient solar cells, bright light emitting devices, quantum sources of light. Quantifying heterogeneity and understanding coupling between dots is critical for these applications. We use double-nanohole optical trapping to size individual dots and correlate to emission energy shifts from quantum confinement. We were able to assemble a second dot in the trap, which allows us to observe the coupling between dots. We observe a systematic red-shift of 1.1$\pm$0.6~meV in the emission wavelength. Theoretical analysis shows that the observed shift is consistent with resonant energy transfer and is unusually large due to moderate-to-large quantum confinement in PQDs. This demonstrates the promise of PQDs for entanglement in quantum information applications. This work enables future in situ control of PQD growth as well as studies of the coupling between small PQD assemblies with quantum information applications in mind.
\end{abstract}
\section{Introduction}
 Perovskite quantum dots (PQDs) show intriguing properties for quantum technologies, such as bright and highly coherent single photon emission~\cite{utzat2019coherent,zhu2022room}, and superfluorescence in ensembles of dots~\cite{raino2018superfluorescence}.  Coupled quantum dots have long been investigated for quantum computing~\cite{burkard1999coupled,loss1998quantum}. Many fabrication strategies have been proposed to couple different types of quantum dots, not only PQDs~\cite{kim2011ultrafast,rodary2019real, koley2021coupled,panfil2019electronic,cui2019colloidal,cassidy2021tuning}. Assembling quantum dots and studying their quantum coupling in solution would greatly simplify such studies. Furthermore, by assembling in solution, it is possible to study the individual dots prior to assembly, and then study the impact of coupling in the near-field, which is not possible with pre-assembled pairs. 
 
 Studying individual dots also allows for probing non-uniformity of the PQDs in solution. Uniformity has been recognized as an important parameter for high-performance applications~\cite{krieg2020monodisperse,tennyson2018cesium,bhattacharyya2019mechanochemical,zhao2019size, cheng2020size,doherty2020performance,brown2021precise}. 
For solar cells, monodisperse PQDs have shown higher conversion efficiencies and open circuit voltages~\cite{lim2021monodisperse}. PQDs are recognized as highly coherent single photon emitters~\cite{utzat2019coherent,zhu2022room}; however, for indistinguishable photons, nearly-identical emitters are desired and this requires a way to select among individual emitters in the ensemble~\cite{santori2002indistinguishable,gschrey2015highly}. Past efforts have focused on ex-situ characterization (e.g., transmission electron microscopy -- TEM) of already synthesized PQDs. Ideally, particle size would be monitored in solution in real time, allowing for in-situ tailoring of growth conditions while preventing degradation from exposure to the environment. 

Here, we use double-nanohole (DNH) optical tweezers to characterize the dispersion of cesium lead bromide (CsPbBr$_3$) PQDs and their coupling in solution. Aperture based optical tweezers have been used to trap quantum dots~\cite{zehtabi2013double}, study their emission (also with two photon excitation)~\cite{jensen2016optical}, and enhance their single photon emission characteristics~\cite{jiang2021single}. We demonstrate that the DNH optical tweezer can be used to determine the sizes of individual PQDs and correlate size with the emission spectra shifts from quantum confinement. We also demonstrate that the DNH tweezer can capture two quantum dots (i.e., assemble them in real-time) and thereby measure the spectral shift that arises from their coupling. Therefore, this platform enables the spectral and size characterization of single and double dots, and most importantly, it achieves this feat in-situ without removing the dots from the solution or requiring electron microscopy that would damage them. 

Multiple physical mechanisms can be responsible for  quantum dot coupling in solution. For quantum dots that are nominally symmetric and do not allow for electron tunneling, F\"orster resonant energy transfer (RET) has been considered as a way to achieve coupling between dots, providing a possible avenue towards quantum information processing~\cite{harankahage2021quantum}. In the past, RET has been studied for PQDs of different sizes, where the longer wavelength emission peak is enhanced due to one-directional energy transfer~\cite{de2016energy}. The bi-directional coupling that arises from just two PQDs that are nominally the same size has not been investigated so far.

\begin{figure}[htbp]
\centerline{\includegraphics[width=160mm]{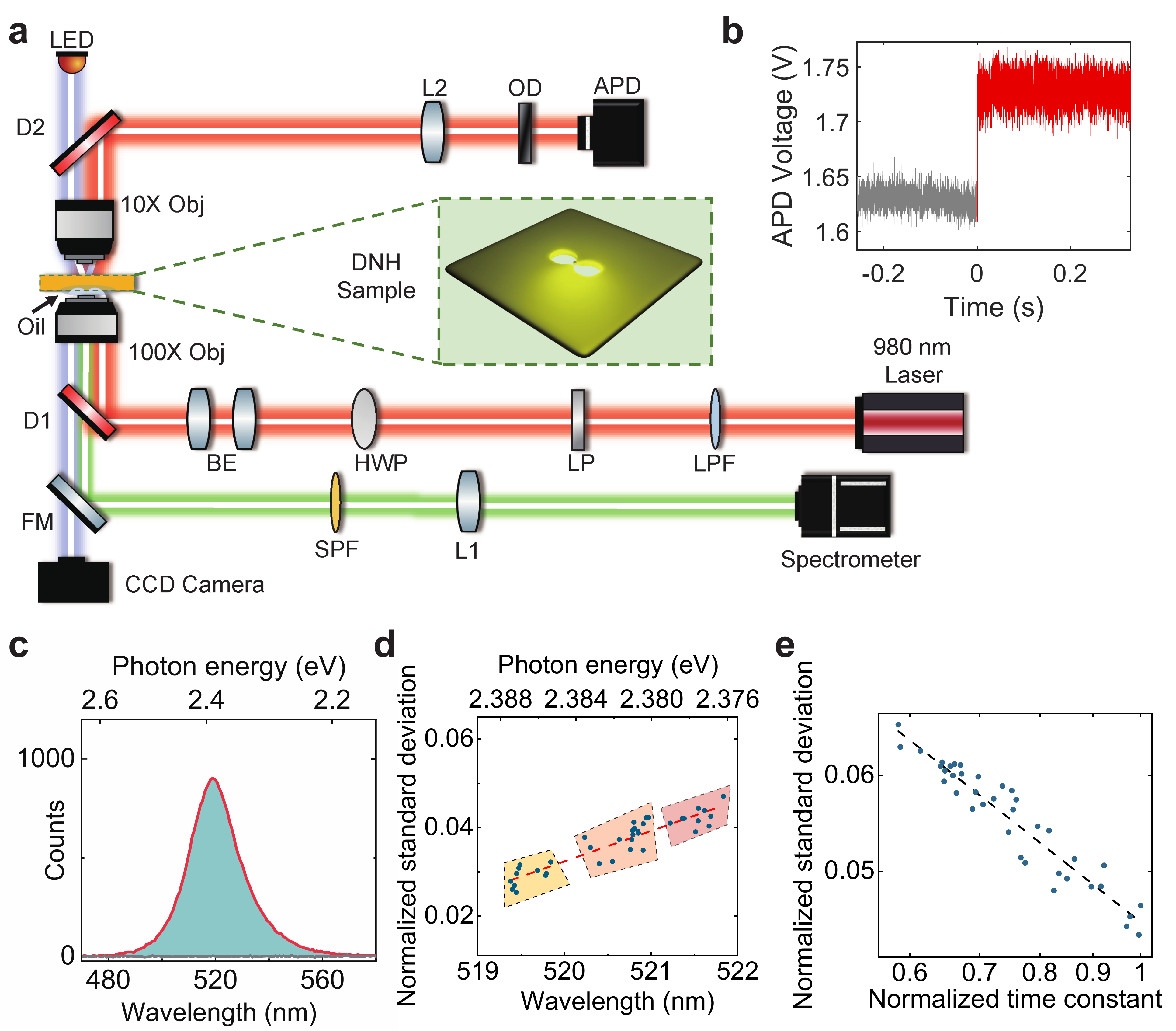}}
\caption{(a) Optical trapping setup: long pass filter (LPF), linear polarizer (LP), half-wave plate (HWP), beam expander (BE), dichroic mirror (D), objective lens (Obj), lens (L), optical density filter (OD), avalanche photodetector (APD), flip mirror (FM), short-pass filter (SPF), charge-coupled device (CCD).
(b) Single trapping event (untrapped: grey, trapped: red). (c) Emission in situ
(untrapped: grey, trapped: red)
. (d) Relation between emission wavelength and standard deviation of trapping laser transmission for individual trapped PQDs. Three groups of the sample were measured. This shows that the standard deviation can be used as an independent measure of the particle size, which correlates well with the quantum confinement induced blue-shift. (e) Standard deviation against autocorrelation time constant ($\tau$) of the single quantum dots trapping events on log-log plot. The slope for the best linear fit is about -0.68.}
\label{fig1}
\end{figure}

\section{Results and discussion}

\subsection{Heterogeneous Particle Sizing}

CsPbBr$_3$ PQDs were synthesized by the ligand-assisted reprecipitation technique~\cite{du2017high}. 
The side-length distribution calculated by averaging over 150 PQDs is 11.8 $\pm$ 1.5 nm . Also shown is the distribution of smallest edge sizes, since the PQDs are expected to align preferentially along their long sides when pushed together by optical forces, and so the separation is best represented by their small side size. We find the small side size to be 10.5 $\pm$ 1.1 nm, as shown in the Supporting Information (SI), Fig. S1; however, this size showed variation of 0.5~nm between separate fabrication runs. We diluted the sample with toluene 20-fold, then placed the solution in a microwell formed by an imaging spacer (Grace Biolabs) on a slide 0 coverslip. A gold film containing DNHs on glass, fabricated using past approaches~\cite{ravindranath2019colloidal}, was placed on top of the microwell to seal. We then placed the sample in an inverted microscope optical tweezer setup adapted from a modular kit (Thorlabs -- OTKB), shown in Fig.~\ref{fig1}a, which has been used previously to study lanthanide nanocrystals~\cite{sharifi2021isolating,alizadehkhaledi2019isolating}. The modified setup included a spectrometer for measuring the spectrum of PQD(s) in the DNH optical tweezer. The excitation of the PQD was obtained by a two-photon process, since the laser wavelength was 980 nm (1.26~eV), and the emission wavelength was 520~nm (2.38~eV). Two photon excitation of colloidal quantum dots has been observed previously in trapping setups~\cite{jauffred2010two, jensen2016optical}. We confirmed the two-photon process by the quadratic power dependence, as outlined in the SI, Figure S4. The observed wavelength is similar to past works for similarly sized dots~\cite{utzat2019coherent}, but longer than smaller perovskite dots~\cite{krieg2020monodisperse}.

A step in the transmission of the laser through the DNH was seen with trapping as shown in Fig. \ref{fig1}b. This step resulted from the dielectric loading of the DNH by the PQD, which made the aperture optically ``bigger". Several works have confirmed that the single step is the result of an individual nanoparticle being trapped, e.g., by using fluorescent particles~\cite{berthelot2014three} or by noting the dynamics when multiple particles are trapped~\cite{chen2012enhanced}.  
The spectrum of the single dot is shown in Fig. \ref{fig1}c, which was recorded after the trapping occurred. There was no emission observed before trapping. 

It was possible to estimate the size of each isolated trapped PQD by analyzing thermal motion induced fluctuations. As demonstrated previously for proteins (but not for PQDs), the autocorrelation time of the thermal motion scales as $\gamma/\kappa$, where $\gamma$ is the drag coefficient (which scales as the cross sectional width in the viscous limit) and $\kappa$ is the optical tweezer stiffness (which scales as the volume of the particle in the dipole limit). Therefore, for radius $r$, $\tau \propto r^{-2} \propto \Omega^{-2/3}$~\cite{kotnala2014quantification,wheaton2015molecular}, where $\Omega$ is the volume. Also, the standard deviation of the light scattering scales linearly with the size~\cite{wheaton2015molecular}. This has been applied to study heterogeneous solutions of proteins~\cite{hacohen2018analysis}. Here we apply it to sizing individual PQDs and correlating the size with the emission spectrum shifts from quantum confinement. 

Fig.~\ref{fig1}d shows the standard deviation of the trapping laser fluctuations correlated with the emission wavelength for individual PQD trapping events. This data was taken from three separate batches, and shows a clear separation in the sizes of these batches. As described above, it is expected that the standard deviation scales linearly with particle size. For small (first-order) variations in size, it is also expected that the wavelength scales linearly with particle size. Therefore, we observe a linear relationship between the standard deviation and the emission wavelength. We considered the autocorrelation time as a separate measure of particle size, as shown in Fig.~\ref{fig1}e. We observed that the relation between the standard deviation and the autocorrelation time had a -0.68 slope on a log-log plot, where a slope of -2/3 is the theoretical prediction~\cite{wheaton2015molecular}. The detailed calculation is shown in the SI. 

The size-dependence of the emission spectra of individual PQDs can be modeled by solving the Schr{\"o}dinger equation under the effective mass approximation~\cite{brus1986electronic}. While spherical particles with infinite barriers allow analytical calculations~\cite{brus1986electronic}, here we also use numerical calculations which allows for cubic particles and finite barrier energies~\cite{yang2020understanding}. Details of the calculation method are provided in the SI. 
The ability to size and spectroscopically characterize individual dots in solution is relevant for quantum applications where we seek to obtain multiple indistinguishable emitters~\cite{utzat2019coherent}. The method can also be used in-situ to optimize solution-based growth~\cite{wu2017monitoring,zhao2019size,cheng2020size,doherty2020performance,brown2021precise}. 

\subsection{Two Quantum Dot Assembly via Trapping (Dimers)}

\begin{figure}[htbp]
\centerline{\includegraphics[width=165mm]{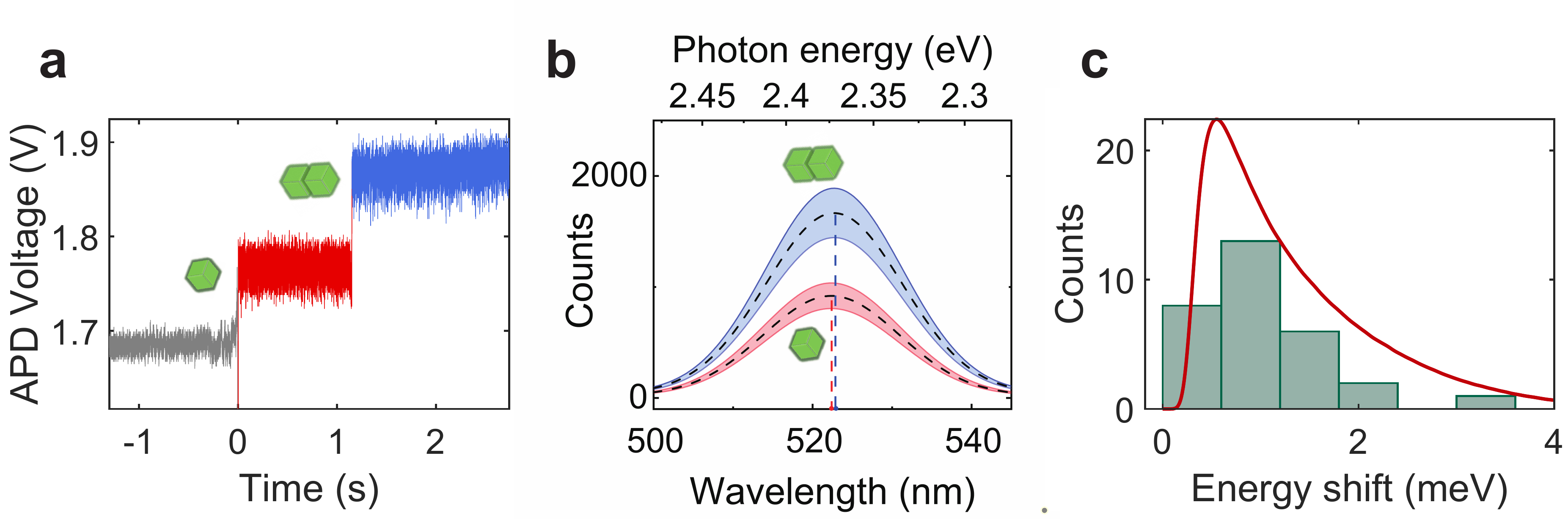}}
\caption{(a) Double trapping event collected by APD (first trapping: red; second trapping: blue). (b) Spectra of single dot trapping and double dot trapping are shown in red and blue. The areas filled with blue and red stand for the experiment data (fitted) at each sampling time. Black dotted lines indicate the average of the varying spectra region (quantum blinking). (c) Experimental (green bars) and simulated (red curve) energy shift upon coupling. Experimental values were observed among 30 different DNHs, each measured at least 6 times, and the simulation represents the energy shift by F\"orster interaction between two dots randomly selected from 1000 dots (normal distribution). 
}
\label{fig2}
\end{figure}

The trapping setup allowed us to measure and characterize the quantum coupling between two PQDs assembled in real time and isolated in the trap.
We observed two steps corresponding to double PQDs trapped subsequently in the same aperture as shown in Fig. \ref{fig2}a. It has been shown in past works on aperture optical tweezers that the co-trapping of two nanoparticles results in a double step profile~\cite{chen2012enhanced}. 
Fig.~\ref{fig2}b shows the observed distribution of emission for single and double PQD trapping averaged over 30 different DNHs, averaging over at least 6 measurements at each DNH (full data is included in the SI, Figure ~\ref{Supp4}).  Fig. ~\ref{fig2}b, shows that the emission intensity approximately doubled when there are two PQDs,  and that there was a systematic spectral red-shift of the emission. Fig.~\ref{fig2}c quantifies the systematic red-shift in the emission observed for the multiple double-dot trapping events. It is relevant to note that there was always a red-shift: if the energy was simply transferred from the smaller dot to the larger one (as is the case for past works on RET~\cite{de2016energy}), we would expect on average a red-shift sometimes and no shift at other times, depending on whether a smaller or a larger dot was trapped first.

Before discussing the physical mechanism that may be responsible for this red-shift, let us describe a simple generic model for the energy shift arising from a generalized coupling potential $V$. Considering two oscillators with energies $E_1$ and $E_2$, the shift from $E_1$ upon coupling can be calculated as~\cite{lovett2003optical, harankahage2021quantum}:
\begin{equation}
    \Delta = \frac{1}{2}\left [ (E_2-E_1)-\sqrt{(E_1+E_2)^2-4(E_1E_2-V^2)}\right ].
    \label{eq:1}
\end{equation}
Fitting $\Delta$ to the observed shift, we find a best fit for $V = 1.1$ meV, as shown in Fig.~\ref{fig2}c.

\subsection{Coupling Mechanisms Between Perovskite Quantum Dots}

We now discuss a possible physical mechanism underlying the observed systematic spectral red-shift. First, we discuss the expected photon coupling between two PQDs, treated as point dipoles.
As shown in the SI, 
the {\it radiative} decay rate from a single dipole emitter
is 
\begin{equation}
    \Gamma_{1}({\bf r}_1)
    = \frac{2 {\bf d}_1 \cdot {\rm Im} \{{\bf G}({\bf r}_1,{\bf r}_1,\omega)\} \cdot {\bf d}_1}
    {\epsilon_0 \hbar},
\end{equation}
for an emitter with dipole moment ${\bf d}_1$ (assumed real)
at position $\mathbf{r}_{1}$, and 
${\bf G}$ is the photon Green function, for any generalized medium.
For coupled dipoles (the second one is with real dipole moment $\mathbf{d}_{2}$ at $\mathbf{r}_{2}$),
the photon exchange has real and imaginary parts, defined through
the incoherent rates of photon transfer:
\begin{equation}
    \Gamma_{12}
    = \frac{2 {\bf d}_1 \cdot {\rm Im}\{ {\bf G}({\bf r}_1,{\bf r}_2,\omega)\} \cdot {\bf d}_2}
    {\epsilon_0 \hbar},
\end{equation}
and a coherent exchange term
\begin{equation}
    \Delta_{12}
    = -\frac{{\bf d}_1 \cdot {\rm Re} \{{\bf G}({\bf r}_1,{\bf r}_2,\omega)\} \cdot {\bf d}_2}    {\epsilon_0 \hbar}.
    \label{eq:Fm}
\end{equation}
The latter gives rise to spectral frequency shifts through photon exchange.

\begin{figure}[hbpt]
    \centering
    \includegraphics[width=0.95\columnwidth]{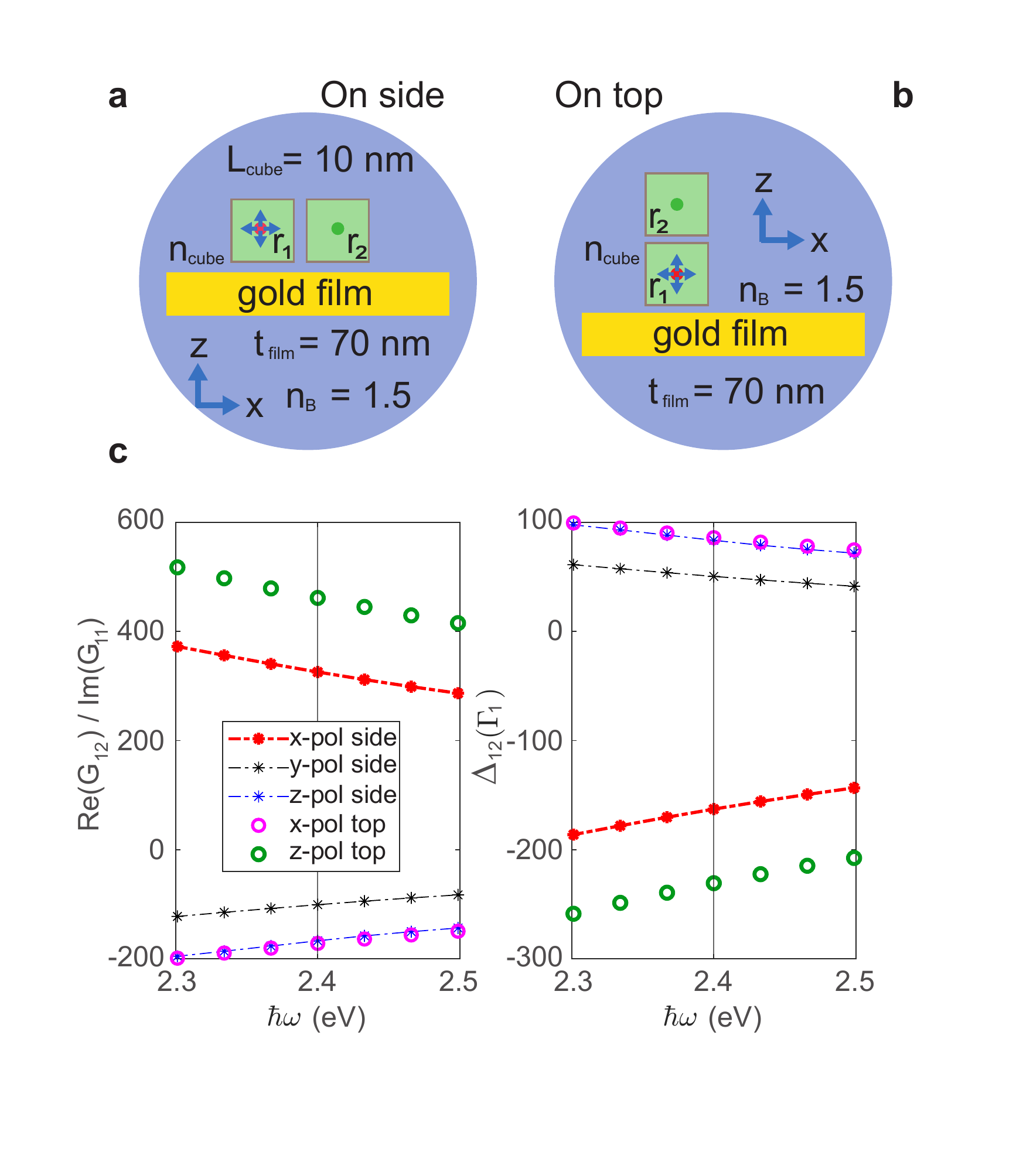}
    \caption{Schematic diagram for PQD dimer with (a) side coupling (left) or (b) top coupling (right). PQDs are shown in green in the two figures. (c) Example calculations for the propagators and dipole-dipole shifts computed from COMSOL, which includes the full scattering geometry. We show different dipole polarizations, and dipole-dipole coupling for PQD cubes that are horizontally stacked or vertically, and separated by 1-nm. All terms include local field corrections, so are normalized by the result including the single PQD cube.}
    \label{fig3}
\end{figure}

When one considers a homogeneous medium (with $\epsilon_b=n_b^2$ using
two dipoles spatially separated by $r_{12}=11~$nm (center to center),
and in the near-field regime, then
the coherent exchange term,
 with $s$-polarized dipoles,
is
\begin{equation}
\hbar\Delta_{12}^{ss}=
\frac{d^2}{4\pi \epsilon_0 \epsilon_b r_{12}^3}
\approx 93\,\hbar\Gamma_1 =
0.25~{\rm meV},
\end{equation}
while for $p$-polarized dipoles, 
\begin{equation}
\hbar \Delta_{12}^{pp}=
-\frac{d^2}{2\pi \epsilon_0 \epsilon_b r_{12}^3}
\approx
-186\,\hbar\Gamma_1 
= -0.50~{\rm meV},
\end{equation}
where we have used
$d=0.72\,{\rm e \cdot nm}$ and $\epsilon_b=1.5^2$, yielding
a nominal radiative decay rate of $\hbar\Gamma_1\approx 2.69~{\rm \mu eV}$ (corresponding lifetime of 250 ps); this estimate scales with $\Gamma_{1,2}$ so could easily be larger.
Note that these analytical rates identically
recover the well known F\"orster coupling terms, and can be described semi-classically or quantum mechanically.
Below,  we will introduce the shorthand notation
${\bf G}_{12} = {\bf G}({\bf r}_1,{\bf r}_2)$.
For identical dipole emitters, the resonance will split by $\pm \hbar\Delta_{12}  \equiv \pm V_F \propto {\rm Re}\{G_{12}\}$, into subradiant and superradiant states, where the latter is optically bright. ($V = V_F$ for RET in Eq.~\eqref{eq:1}). Based on particle size variation giving a variation in the centre-to-centre separations, as well as expected dipole moment variation (smaller dots have smaller dipole moments~\cite{Becker2018}), we expect the variation in $V_F$ to be within 50\% of the reported value.

In the SI, we
describe how these radiative decay and exchange rates are influenced 
by local field corrections and interactions with the metal film, from both an analytical perspective as well as using full numerical simulations in COMSOL. As an example of the latter calculations, here we consider
two PQD cubes placed $1$~nm above a gold film, either
horizontally coupled or vertically coupled, as
shown in Figs.~\ref{fig3}a and b. The PQD cubes are separated by 1-nm (gap size), and various dipole polarizations are considered. The corresponding Green function propagators and dipole-dipole coupling rates (in units of the radiative decay rate) are 
shown in Fig.~\ref{fig3}c.
Overall we predict that at $\hbar\omega=2.4$~eV, maximum
photon exchange rates of around $-231\hbar\Gamma_1$
are possible, which we estimate to be around 
$-0.62$~meV (for $z$-polarized dipoles with top coupling). 

After initial calculations at a $1$~nm gap size, we obtained a more accurate measurement of the gap size from high-resolution TEM as being $1.4$~nm. This has a small ($10\%$) impact on the theoretical values (Fig. S14). The typical gap for oleic acid PQDs in TEM imaging is between 1 and 2 nm~\cite{raino2018superfluorescence,protesescu2015nanocrystals,utzat2019coherent,du2017high,li2018excitons}. The optical force tends to squeeze the particles together and minimize the gap size in solution, and so we expect this is similar to what was observed by HR-TEM.

It is also important to note that the expected exciton Bohr radius is much smaller than the size of the
PQDs, which gives rise to a giant oscillator strength for an optical transition~\cite{Becker2018,nair1997theory}.
This is caused by a correlated exciton wave function
that affects both the radiative decay and the dipole-dipole coupling rates. We discuss and estimate this enhancement in the SI, which we expect to be about a factor of 5 bigger than PQDs in the strong confinement regime. This is significant, even in the presence of local field reductions.
Also in the SI, we discuss several other possible coupling mechanisms including Dexter coupling, electronic tunneling, exciton tunneling for fused dots, and possible monopole-monopole interactions. 

\section{Discussion}
\subsection{In-Situ Real-Time Size-Spectral Characterization}

In the present work, we have demonstrated the use of an optical tweezer platform to characterize multiple individual PQDs in solution. 
In past works, the standard approach for monitoring PDQ synthesis has been to follow up PDQ growth with characterization via TEM or luminescence studies. Here we show that we can accurately determine the size and the spectral response without the need for TEM. This may allow for fine-tuning or improving growth conditions in-situ. For example, it is possible to envision integrating the present trapping setup with a flow-cell 
~\cite{zehtabi2013double}
for real-time monitoring and/or modifying of growth conditions~\cite{wu2017monitoring}.

\subsection{Quantum Dot Coupling}

We believe that RET is the dominant coupling mechanism between the two PQDs.
We do not think the observation can be explained by other mechanisms, like Dexter coupling and quantum hybridization, because the barrier height is too large and the gap too wide to allow substantial tunneling~\cite{xue2019small}. RET has been observed in ensembles of dissimilar PQDs~\cite{de2016energy}; however, that exchange is uni-directional (from the higher energy to the lower energy PQD). Here, since the dots are nominally the same size, the exchange is bidirectional. For quantum applications, RET has been proposed theoretically as a mechanism to achieve quantum computing via colloidal quantum dots~\cite{harankahage2021quantum}. Here we estimate the magnitude of the RET induced shift. The dipole moment is estimated from previously reported values of the emission lifetime. We further note that the PQDs studied here are large relative to the exciton Bohr radius, and are in the intermediate to weak confinement regime. This enhances the strength of the RET interaction by 5 times with respect to strongly confined dots, which is a particularly relevant finding of this work.

We estimate from a dipole model that the shift is at least two orders of magnitude larger than the radiative lifetime. We have also looked at distributed wavefunctions (beyond the local dipole model), and full COMSOL simulations including the metal surface nearby in aperture, and these only provide small corrections to the value reported above (as described in the SI). Since this value is comparable to what we observed in the experiments, RET is a plausible explanation for the observed shift. In the future, we aim to modify the ligands and the dot size to investigate their impact on the RET, since there is a strong size-dependence of this effect.

\section{Conclusions}

We have demonstrated the ability to characterize the size and coupling between individual and dimer PQDs in solution, and simultaneously observe their emission spectra using nanoplasmonic tweezers. This is a powerful tool to quantify sample heterogeneity in size and emission in solution, without requiring expensive and damaging techniques. The approach may be extended to isolate identical dots (both in terms of size and spectral emission), which is a pathway to achieving indistiguishable quantum emitters. The approach may also be extended to monitor and modify growth in real time.

We quantified the systematic red-shift for coupled PQDs in solution of 1.1 $\pm $ 0.6 meV and we argue that this is likely the result of resonant energy transfer. The magnitude of the RET is unusually large when compared with the strongly confined quantum dots that have been explored in the past, which is intriguing because RET has been proposed as a mechanism to obtain entanglement for quantum information processing applications~\cite{harankahage2021quantum}. 

In the future, we aim to consider the impact of temperature on the observed spectral shifts, which will require adding cyrogenic capability to our setup so that the temperature can be lowered after trapping is achieved. To first order, RET does not depend on temperature~\cite{andrews2011resonance,zhang2020temperature}, however, more advanced  electron-phonon theories predict a renormalization of $V_F$
as a function of temperature~\cite{nazir2005anticrossings,nazir2009correlation}. We present only a simple temperature independent model here, as a first order quantification of the RET effect. 
It may be possible to modify the setup to allow for cryogenic cooling after trapping is achieved, and for time-resolved ultra-fast probing of the trapped dots. Combined, these advances will allow for further exploration of the coupling between these quantum emitters. The DNH optical tweezer is an interesting platform for studying regimes of coupling for small assemblies of 2 or more PQDs (however larger apertures would be preferred for multiple dots). In this manner, it may be possible to explore the full range of interactions from the single dot, coupled dots, several dots, to large cluster superfluorescence ensemble measurements~\cite{raino2018superfluorescence}; thereby transitioning from the nanoscopic to mesocopic regimes.

\begin{acknowledgement}

The authors thank the NSERC CREATE in Quantum Computing program and the fabrication facilities of CAMTEC. We also acknowledge NSERC for funding through the Discovery Grants program, the Canadian Foundation for Innovation (CFI) for computational infrastructure funding through  the Innovation Fund, and CMC Microsystems for the provision of COMSOL Multiphysics.

\end{acknowledgement}

\section{Declarations}
The authors declare no competing interests.

\section{Contributions}
H.Z. fabricated the DNHs, performed the trapping experiments, analyzed the data and did the FDTD simulation. P.M. and V.Y. made the PQD samples. M.I.S. advised on PQD synthesis and characterization. C.C. and A.B. were responsible for TEM imaging. B.H. and I.P. contributed to simulation of bandgap and emission energy as a function of single PQD size. J.R. and S.H. contributed to the theory and simulations of coupled quantum dots. R.G. advised on the experiments and analysis. All authors contributed to writing the manuscript.

\begin{suppinfo}
Fabrication details, detailed optical trapping description, bandgap and emission characteristics of PQDs, and theory and simulations of coupled quantum dots.\\

\end{suppinfo}


\clearpage
\section*{Supporting Information}
\setcounter{figure}{0}
\renewcommand{\figurename}{Fig.}
\renewcommand{\thefigure}{S\arabic{figure}}

In this Supporting Information document, we provide further details on the following: \\
 
\noindent S-I Fabrication. \\
S-II Optical trapping setup and trapping process.\\
S-III Bandgap and emission energy as a function of single PQD size.\\
S-IV Theory and simulations of coupled quantum dots.

\section{Fabrication}

\noindent \textbf{Materials used for CsPbBr3 PQDs synthesis}

 Cesium bromide (99.9\%), lead bromide ($>$98\%), oleylamine (technical grade, 70\%), oleic acid (technical grade, 90\%), N,N-dimethylformamide (anhydrous, 99.8\%) and toluene (anhydrous, 99.8\%). All chemicals were purchased from Millipore-Sigma.

\vspace{0.2cm}
\noindent \textbf{CsPbBr3 PQDs synthesis method}

\begin{figure}[htbp]
\centerline{\includegraphics[width=180mm]{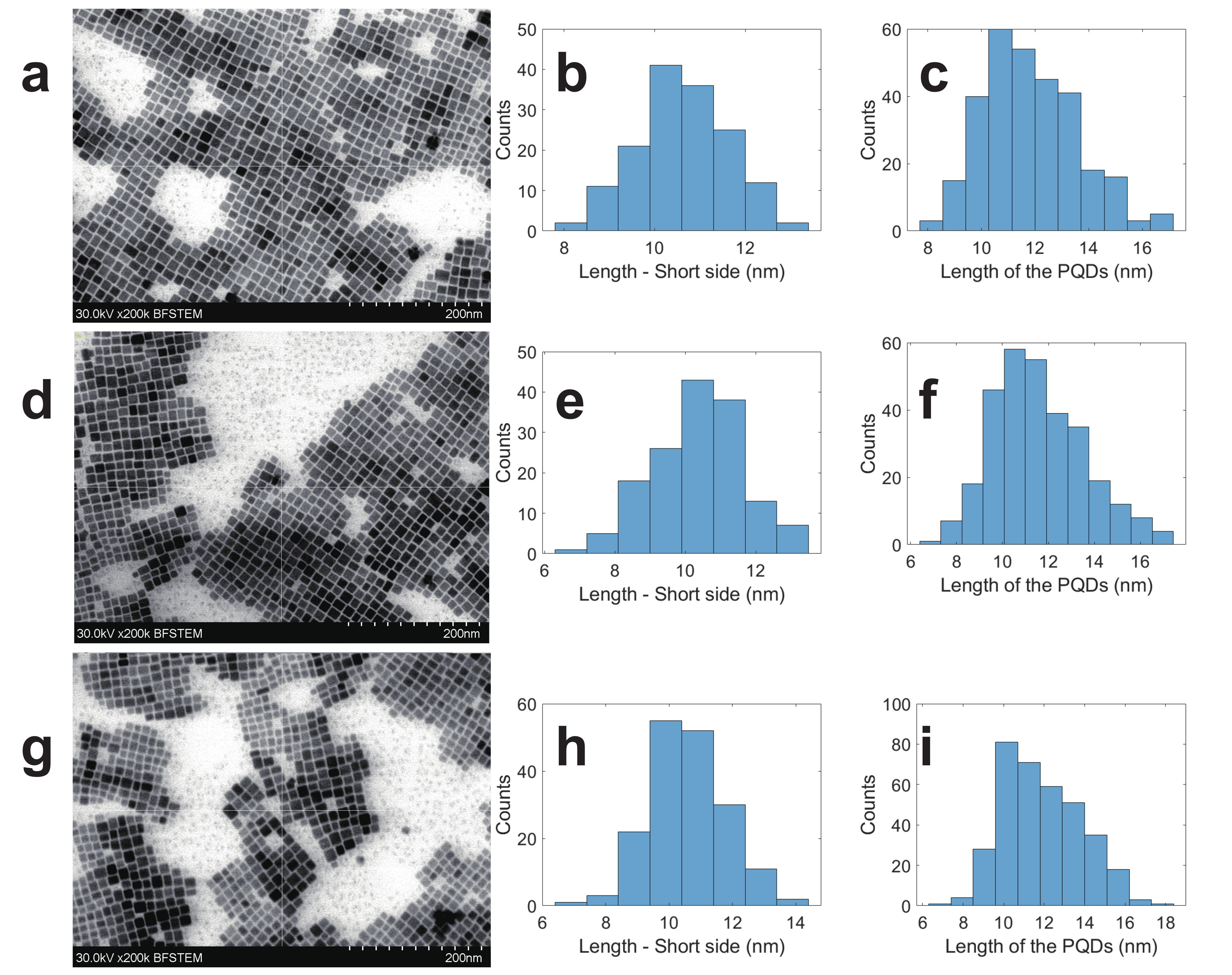}}
\caption{(a-i) STEM images of the CsPbBr$_3$ sample and the side-length distribution of PQDs.}
\label{STEM}
\end{figure}

\begin{figure}[htbp]
\centerline{\includegraphics[width=160mm]{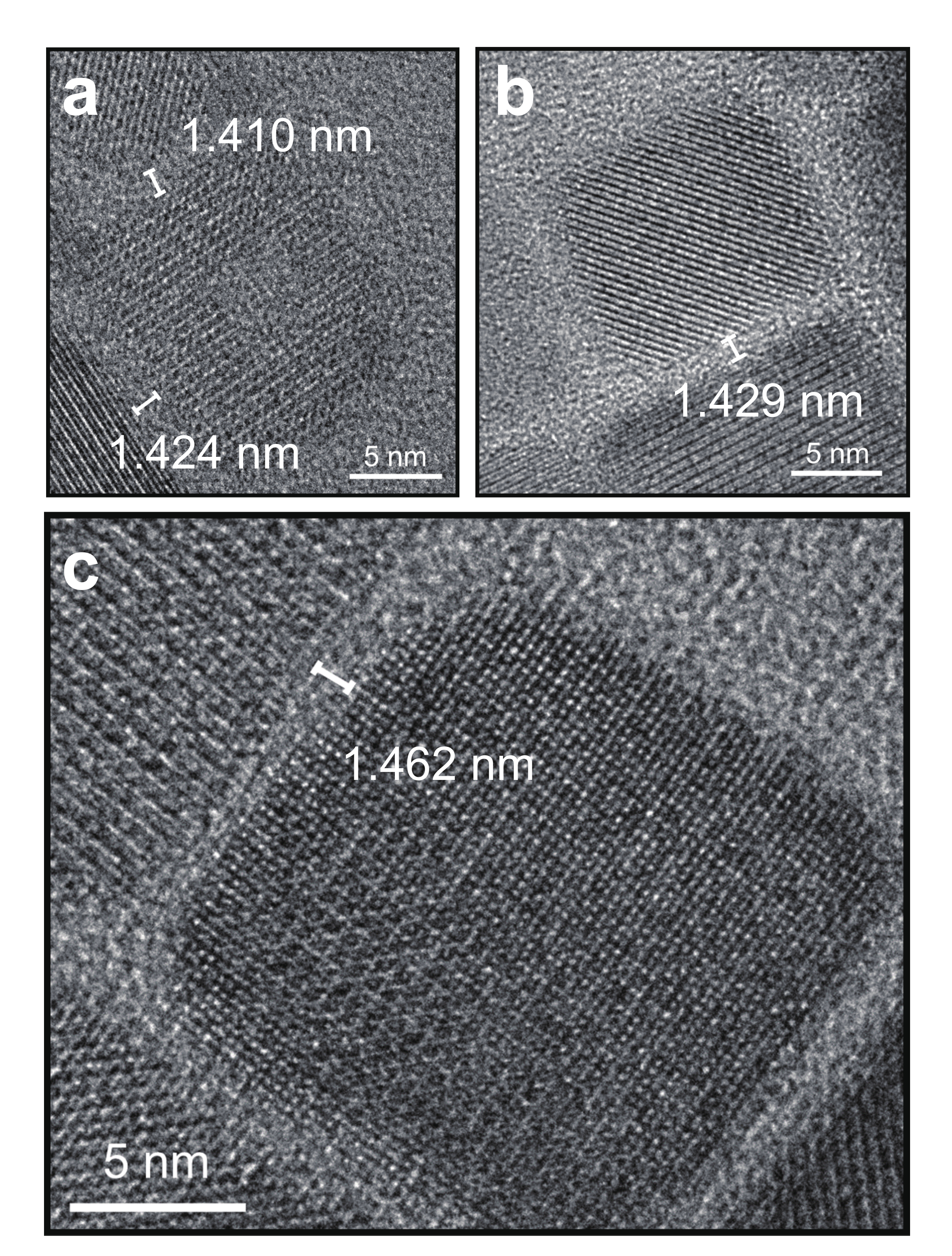}}
\caption{(a-c) High resolution bright field TEM image of PQD with ~1.4 nm ligand.}
\label{TEM}
\end{figure}

 4.25 mg of cesium bromide and 14.68 mg of lead bromide were dissolved in 1 ml of N,N-dimethylformamide. To this solution, 5 $\mu$l of oleylamine and 125 $\mu$l of oleic acid were added. This forms the PQD precursor solution. In another vial, 2.5 ml of toluene was stirred vigorously at 1500 rpm. 250 $\mu$l of PQD precursor solution was dropped quicky into toluene under stirring. The solution color immediately changed to green indicating the formation of CsPbBr$_3$ PQDs. This solution was filtered using 0.2 $\mu$m polytetrafluoroethylene syringe filter. The solution was then used for optical trapping experiments.
Fig.~\ref{STEM} shows the bright field scanning transmission electron microscope (STEM) images of the as fabricated PQDs and their size distribution. Fig.~\ref{TEM} gives high resolution bright field (BF-) TEM images, collected in an in-focus condition with near parallel illumination in an aberration corrected Hitachi HF-3300 TEM at 200 kV. Here the cube-like particles aligned with the beam direction. This allows direct measurements of the interparticle spacing to be taken from the BF-TEM images, measuring the width the region where no lattice fringes were observed. This shows the ligand length of the PQD to be around 1.4 nm. The images presented in Fig.~\ref{TEM} show larger particles due to working at the periphery of the deposited region. For measuring particle size, STEM images with a large field of view were used for representative statistics.

\vspace{0.2cm}
\noindent \textbf{DNH sample fabrication}

 The DNH apertures were made by using colloidal lithography, as we did in our past works~\cite{ravindranath2019colloidal}. Microscope slides were cut and cleaned using the ultrasonic box for 10 min in an ethanol bath. Then, the slides were cleaned by using oxygen plasma for 15 min. 300 nm 0.01$\%$ w/v polystyrene spheres diluted in ethanol were drop-coated on the microscope slides thinly and evenly. After the solution evaporated, the prepared slides were plasma etched again (135~s) to reduce the size of the polystyrene spheres. A 5 nm titanium adhesive layer and a 70 nm gold layer were sputter coated (MANTIS sputtering system). The sputtered samples were sonicated for 10 min in an ethanol bath to remove the polystyrene spheres.

~\\

\section{Optical trapping setup and trapping process}

 The setup in Fig. 1 of the main manuscript consists of a 980 nm continuous-wave laser (JDS Uniphase SDLO-27-7552-160-LD). The optical path was collimated, polarized, filtered and expanded until focusing on the sample with a 100$ \times $ oil immersion microscope objective (numerical aperture = 1.25). The 980 nm beam was used for both trapping and excite the PQDs. The setup collected the transmission light using a 10$ \times $ microscope objective from the sample, where the sample placed on the three-axis sample stage with piezoelectric adjustment. The transmission signal was collected and measured by an avalanche photo-detector (Thorlabs APD120A). A half-wave plate (HWP) and a linear polarizer (LP) were adopted to set the polarization of the beam. A 750 nm lowpass filter (Thorlabs FES0750) to reduce the beam intensity from the 980 nm laser. A fiber connected with the spectrometer (Ocean Optics QE65000) collected the emission light from the sample. The flip mirror was set to change the optical path to collect the image from the sample by using the CCD camera. The gold DNH was attached to a coverslip with an adhesive spacer.


\vspace{0.2cm}
\clearpage
\noindent\textbf{Bulk solution characterization}
\begin{figure}[htbp]
\centerline{\includegraphics[width=140mm]{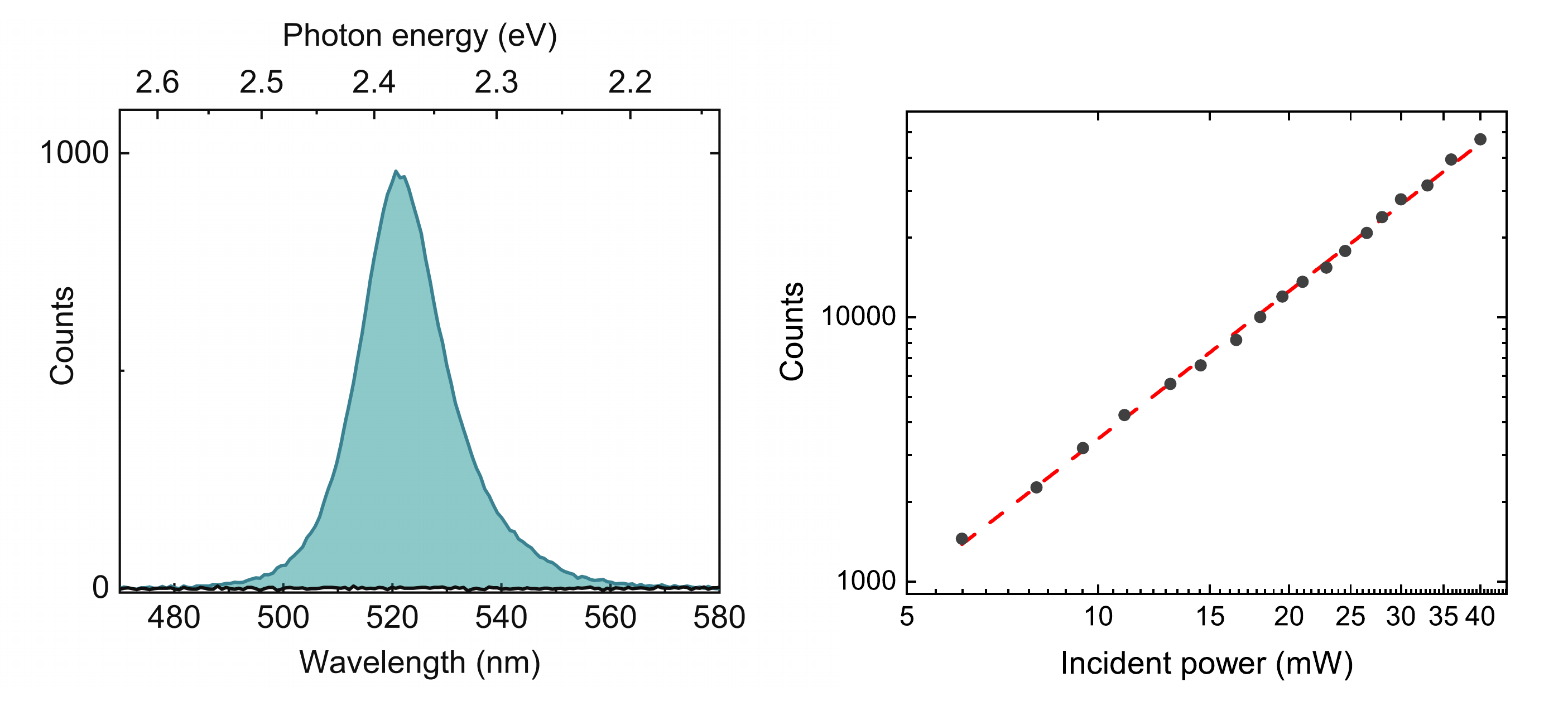}}
\caption{(a) The spectra of the bulk solution. (b) The power dependence of the bulk solution on log-log plot. The slope with best linear fit is around 1.85.}
\label{Supp2}
\end{figure}
 Figure~\ref{Supp2} shows the emission spectrum and the power dependence of emission for an ensemble of PQDs in solution (not trapped in DNH). 

\vspace{0.2cm}

\noindent\textbf{Power dependence}

\begin{figure}[htbp]
\centerline{\includegraphics[width=140mm]{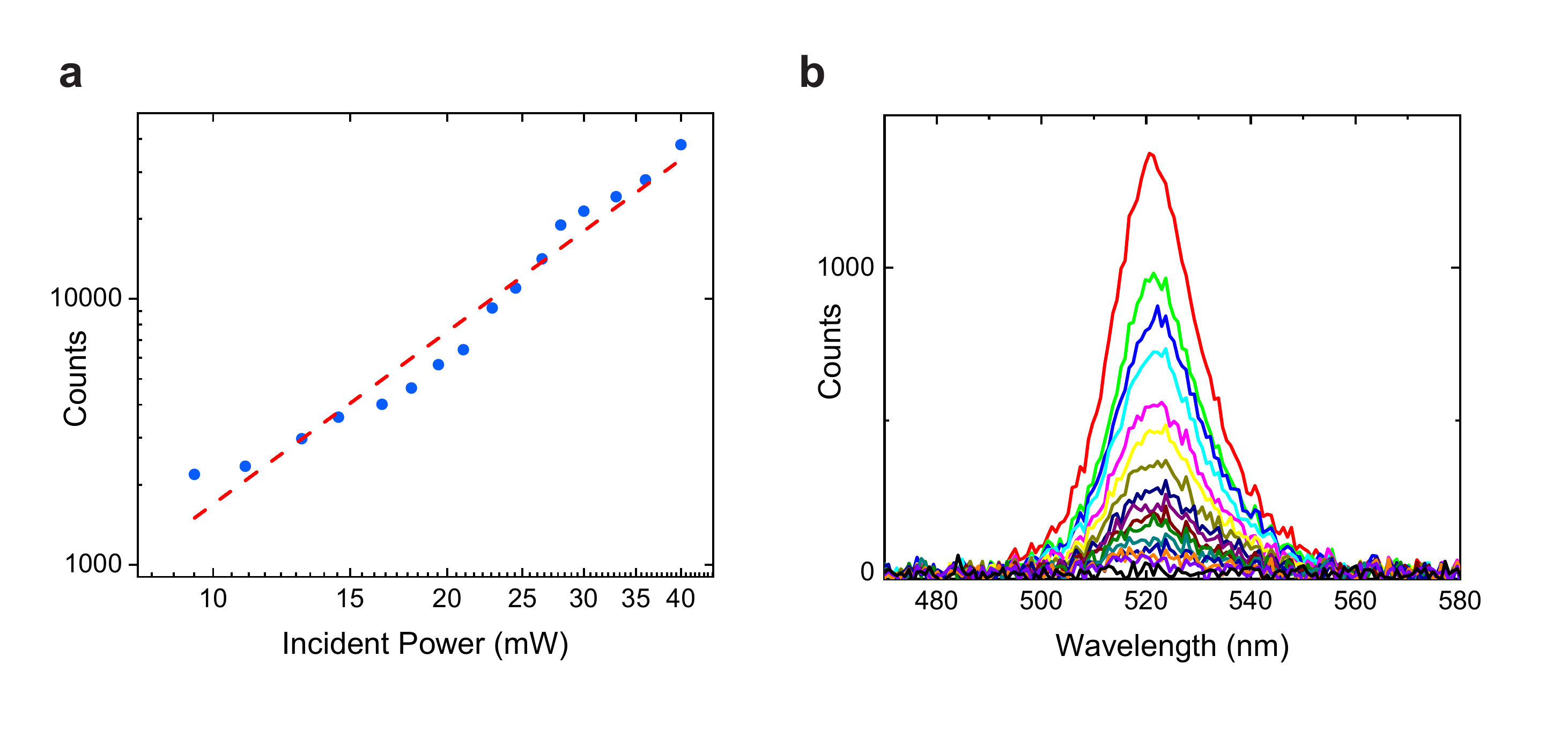}}
\caption{(a) The power dependence of trapping the single quantum dots on log-log plot. The slope with best linear fit is around 2.15. (b) The emission spectra under different incident powers.  }
\label{Supp3}
\end{figure}

Figure \ref{Supp3}a shows the power dependence of the single PQD trapping event. The quadratic power dependence is shown on the log-log plot. The emission counts were integrated from the collected emission peak from 500 nm to 545 nm. Figure \ref{Supp3}b shows that there is little influence with the emission spectra the by increasing of the input power from 8-40 mW in front of the objective lens.

\vspace{0.2cm}
\noindent\textbf{Quantitative analysis by using the thermal motion characteristics}

Analyzing the autocorrelation of the time series signal, an exponential decay time ($\tau$) can be found, which is a measure of the trapping stiffness used in the optical trapping~\cite{rohrbach2005stiffness,kotnala2014quantification}. 
The force on the trapped particle,
\begin{equation}
\label{eq1}
F =m\ddot{x} = -\kappa x-\gamma\dot{x}+F_{L},
\end{equation}
where $\gamma$ is the Stokes' drag, $\kappa$ is stiffness and $F_{l}$ is the Langevin term that accounts for Brownian motion. If we assume $F= 0$ and ignore the noise term in Eq.~\eqref{eq1}, we can obtain $\tau$:
\begin{equation}
\label{eq2}
\tau = \frac{\gamma}{\kappa}
\end{equation}
Since $\kappa \propto \Omega$ and $\gamma \propto \frac{1}{r}$, where $\Omega$ is the volume and $r$ is the radius of the nano-particle, then
\begin{equation}
\label{eq3}
\tau\propto \frac{1}{r^{2}} \rightarrow \tau \propto \frac{1}{\Omega^{\frac{2}{3}}}.
\end{equation}

Using Eqs~\eqref{eq2}-\eqref{eq3}, we can find the relationship between $\tau$ and the volume is $-\frac{2}{3}$. Also, we  then find $F\propto \Omega\propto M$, where $M$ is the mass of the particle. The standard deviation would have a linear relationship with the mass, which can be estimated by the emission wavelength for PQDs with quantum confinement.
\vspace{1cm}

\noindent\textbf{The photoluminescence energy distributions of trapping single and double dots}

\begin{figure}[htbp]
\centerline{\includegraphics[width=160mm]{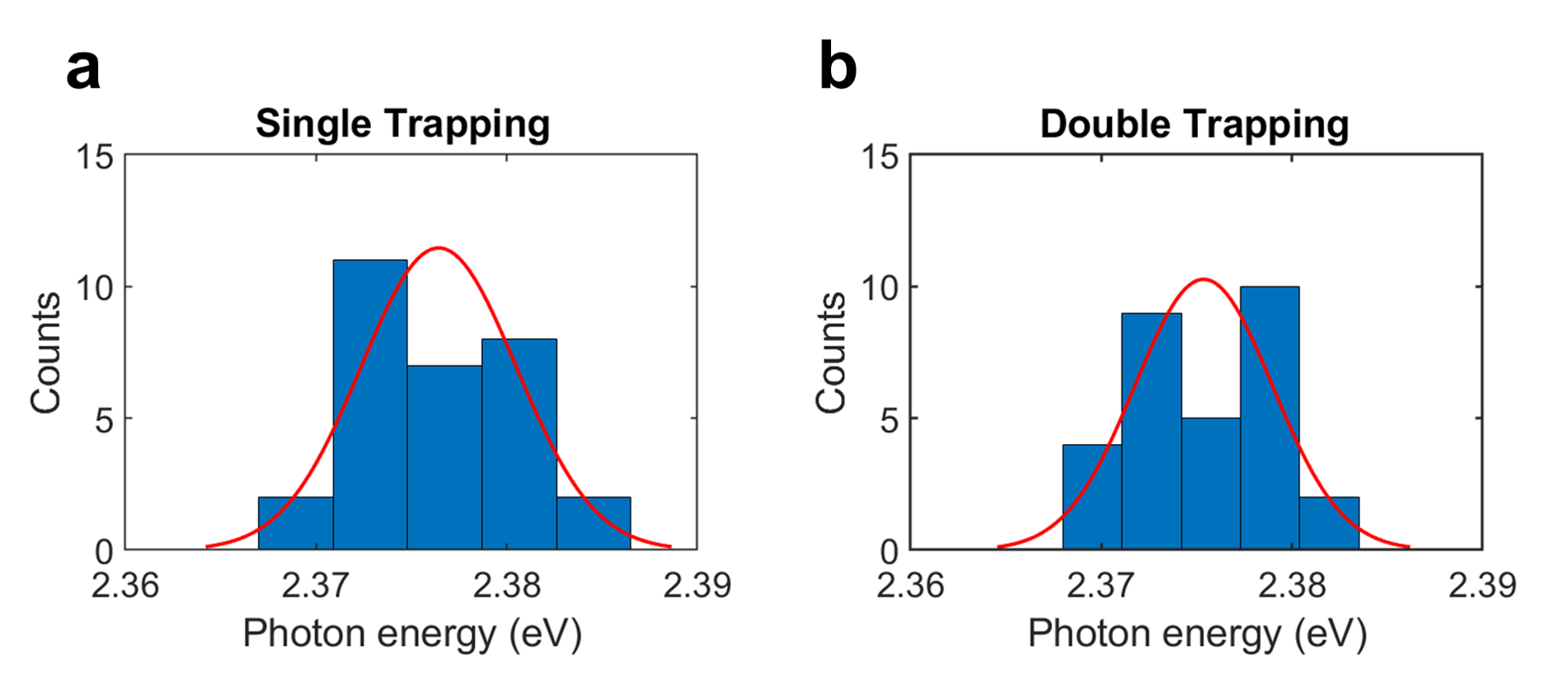}}
\caption{The PL energy distribution of trapping (a) single dot and (b) double dots. The standard deviation of each case is around 4.1 and 3.6 meV.}
\label{distribution_energy}
\end{figure}

Figure~\ref{distribution_energy} a and b shows the photoluminescence energy distributions of trapping single and double dots. The standard deviation of each case is around 4.1 and 3.6 meV.

\vspace{0.2cm}

\noindent\textbf{Reproducible spectra from different DNHs}

\begin{figure}[htbp]
\centerline{\includegraphics[width=170mm]{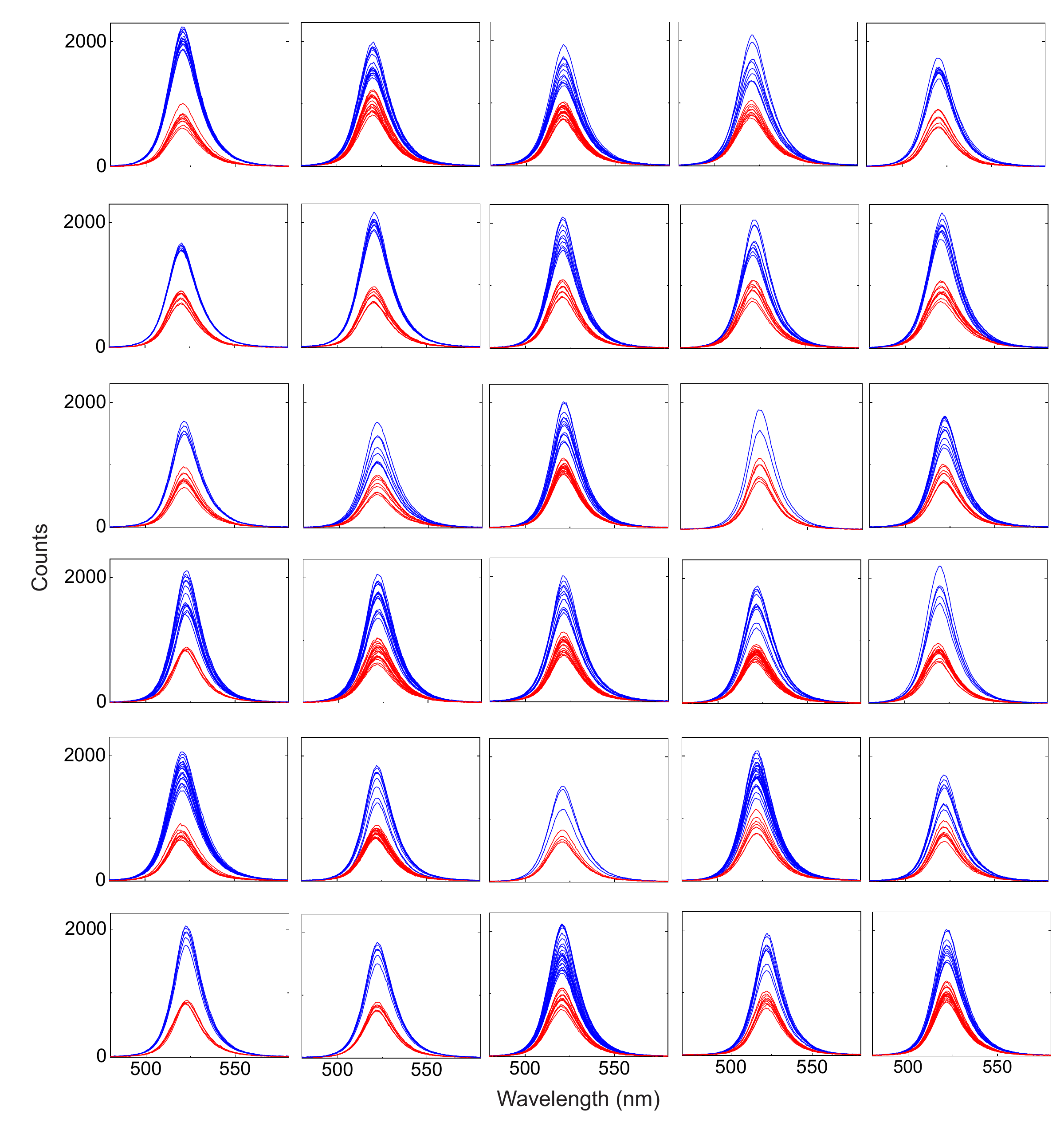}}
\caption{The single and double trapping spectra for different 30 DNHs. Every DNH was measured at least 6 times for reproducible single - double trapping event, respectively.}
\label{Supp4}
\end{figure}

 Figure \ref{Supp4} shows the repeatibility of the results for thirty different DNHs on the same gold sample, with at least 6 trapping events for each DNH. Polynomial fitting is used to find the peak for analysis (concatenation fitting with multiple curves). During the experiment, we observed blinking, which can be seen from the fluctuations in the intensity of the spectra.

\clearpage


\section{Bandgap and emission energy as a function of single PQD size}

\begin{figure}[htbp]
\centerline{\includegraphics[width=120mm]{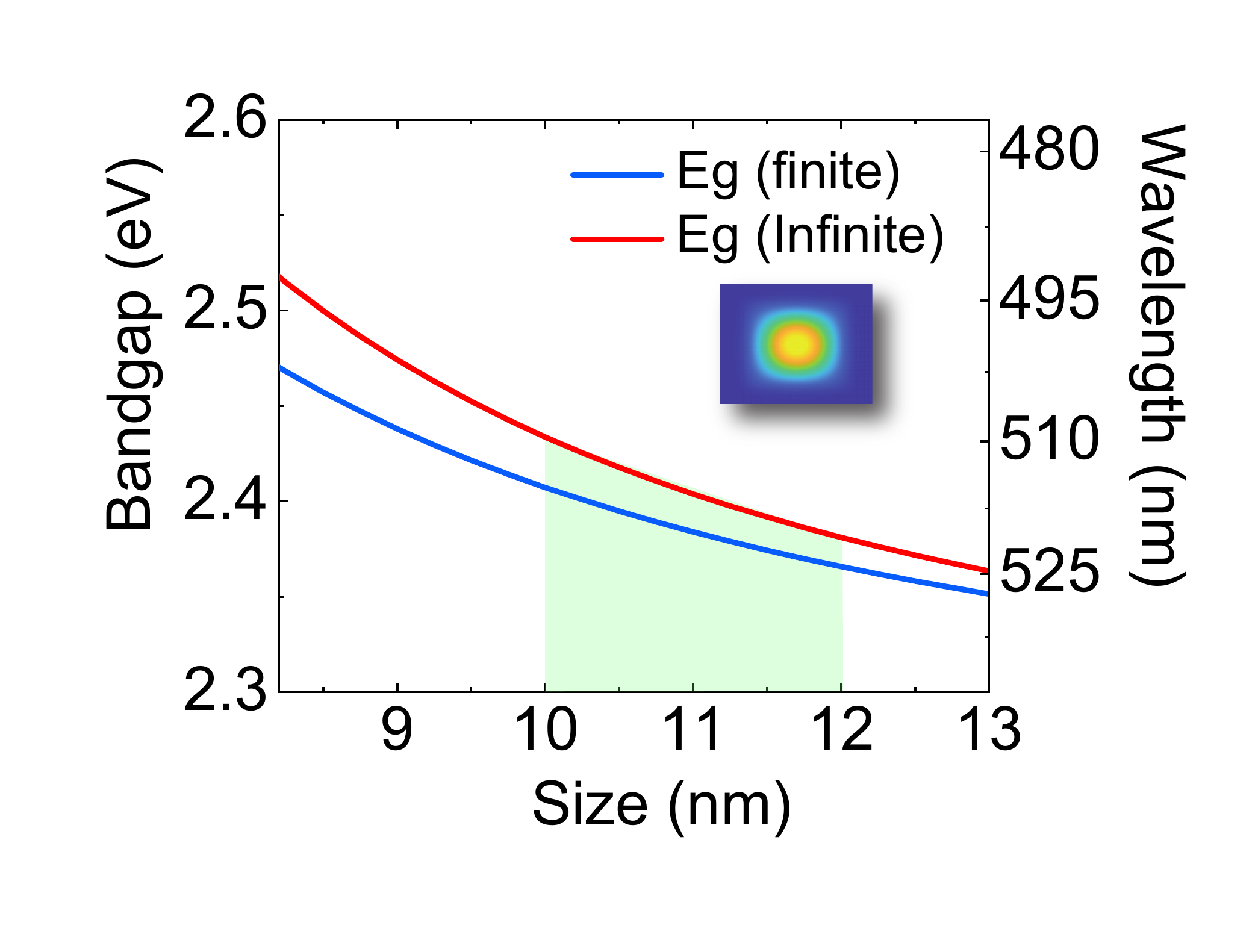}}
\caption{Calculations of size effect for 2 separate cases: Cubic well with infinite and finite barriers. Green highlights the approximate size range observed experimentally. Inset shows the wavefunction distribution for the finite case.}
\label{singleEB}
\end{figure}

We assessed the quantum confinement using an effective mass Schr\"odinger equation model using a Matlab-based finite-element approach, as shown in Fig.~\ref{singleEB} and considering the exciton binding under a weak confinement assumption~\cite{Becker2018}. The simulations used $m_e^* = 0.134m_0$ and $m_h^* = 0.128m_0$~\cite{Becker2018}, with a bulk bandgap of 2.30 eV for cubic CsPbBr$_3$~\cite{ezzeldien2021electronic}, $\epsilon_{QD} = 4.8$~\cite{Becker2018}. 
The barrier height was chosen to either be infinite or to match the band offset between CsPbBr$_3$ QDs and oleic acid ligands (3.14~eV for electrons and 0.79~eV for holes)~\cite{liu2020ligand}. The dispersion found for this model was less than that seen in experiment but it did match with previous works~\cite{Becker2018, protesescu2015nanocrystals}.

\vspace{1.5cm}
\noindent\textbf{Quantum simulation for double QDs }

With the barrier widths seen in TEM imaging (similar to the oleic acid ligand length of $\sim$2~nm) and the barrier heights used above for single dots~\cite{liu2020ligand}, we do not expect significant wavefunction overlap between two adjacent dots. To confirm this, we used effective mass the Schr\"odinger model as above (neglecting the Coulomb potential) and confirmed that there was minimal delocalization for separations greater than about 1~nm.

\vspace{0.6cm}

\section{Theory and simulations of coupled quantum dots}

\noindent\textbf{Photon exchange from coupled dipoles in a general medium  }

The radiative decay rate from a single dipole emitter
is defined from
\begin{equation}
    \Gamma_{1}({\bf r}_1)
    = \frac{2 {\bf d}_1 \cdot {\rm Im}\{ {\bf G}({\bf r}_1,{\bf r}_1,\omega)\} \cdot {\bf d}_1}
    {\epsilon_0 \hbar},
\end{equation}
for an emitter with dipole moment ${\bf d}_1$ (assumed real)
at position $\mathbf{r}_{1}$, and 
${\bf G}$ is the photon Green function.
For coupled dipoles (the second one is with real dipole moment $\mathbf{d}_{2}$ at $\mathbf{r}_{2}$),
the photon exchange mechanism is connected to real and imaginary parts of the Green function, defined from
incoherent rates of photon transfer:
\begin{equation}
    \Gamma_{12}
    = \frac{2 {\bf d}_1 \cdot {\rm Im} \{{\bf G}({\bf r}_1,{\bf r}_2,\omega)\} \cdot {\bf d}_2}
    {\epsilon_0 \hbar},
\end{equation}
and a coherent exchange term
\begin{equation}
    \Delta_{12}
    = -\frac{{\bf d}_1 \cdot {\rm Re}\{ {\bf G}({\bf r}_1,{\bf r}_2,\omega)\} \cdot {\bf d}_2}
    {\epsilon_0 \hbar},
    \label{eq:F}
\end{equation}
where the latter has the same origin
as F\"orster coupling~\cite{thomas}. Specifically,
when one considers a homogeneous medium, the 
near-feld electrostatic part identically
recovers the F\"orster coupling term.
Below, for ease of notation, we will introduce the shorthand notation
${\bf G}_{12} = {\bf G}({\bf r}_1,{\bf r}_2)$.
For identical dipole emitters, the resonance will split by $\pm \hbar\Delta_{12}  \equiv \pm V_F$, into subradiant and superradiant states.
In a Markov approximation, these radiative decay rates are usually evaluated on resonance, which also yield agreement with a Fermi's golden rule approach.

The Green function $\mathbf{G}(\mathbf{r},\mathbf{r}_0,\omega)$,  describes  light propagation from a source point $\mathbf{r}_0$ to $\mathbf{r}$ and is formally defined via the Helmholtz equation 
\begin{equation}
     \boldsymbol{\nabla}\times\boldsymbol{\nabla}\times\mathbf{G}(\mathbf{r},\mathbf{r}_0,\omega)-k_0^2\epsilon(\mathbf{r},\omega)\mathbf{G}(\mathbf{r},\mathbf{r}_0,\omega)=k_0^2\delta(\mathbf{r}-\mathbf{r}_{0}),\label{eq: GreenHelmholtz}
\end{equation}
together with suitable radiation conditions. 

\begin{figure}[hbpt]
  \centering
  \includegraphics[width=0.65\columnwidth]{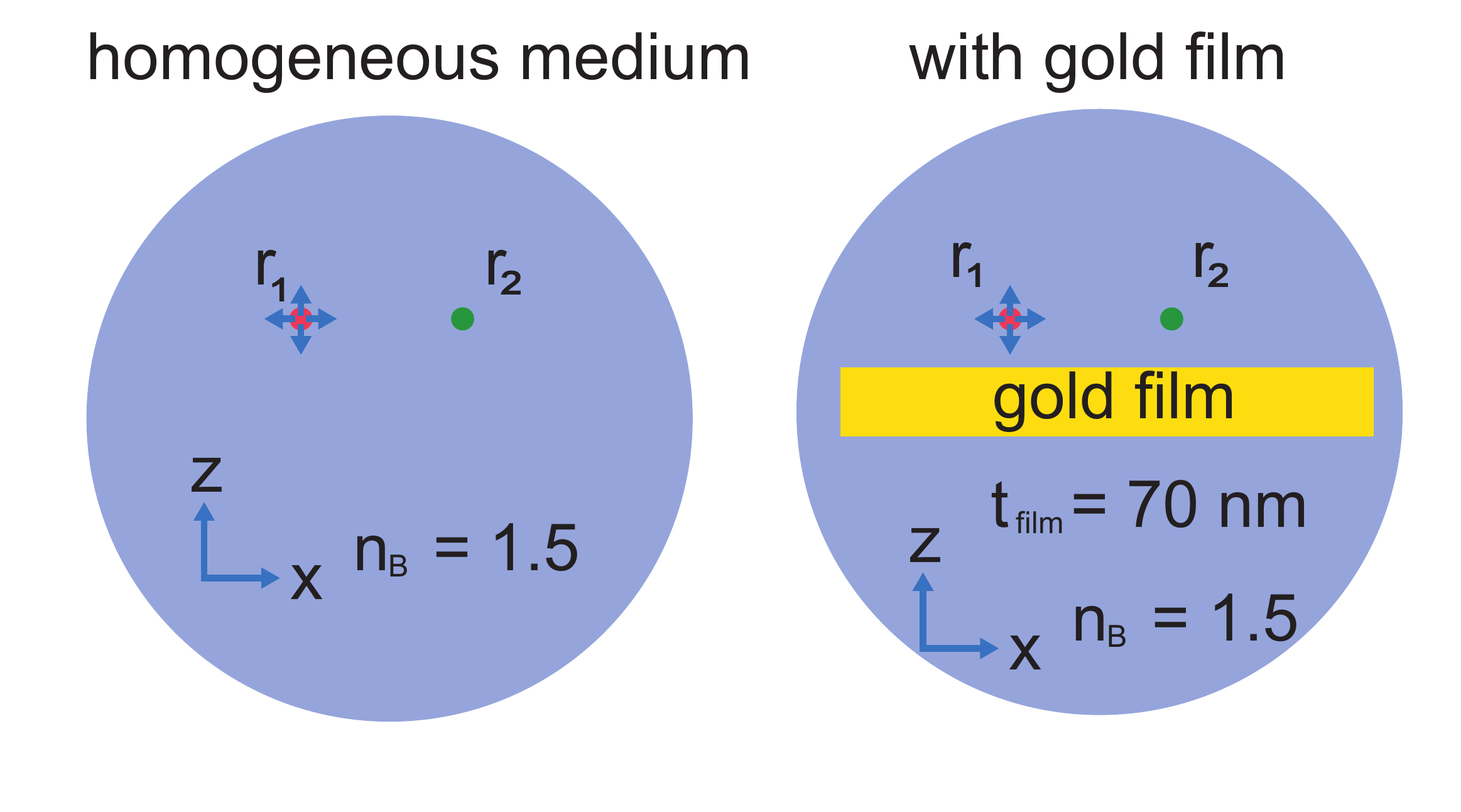}
  \caption{Schematic of two coupled dipoles in homogeneous medium (left side) and above a gold film (right side). They are $11$~nm separated in x direction. With gold film, they are $6$~nm away from the film surface in $z$ direction. The dipoles could be set as x-polarized and z-polarized.
  }\label{fig:film}
\end{figure}

In a background medium
with permittivity $\epsilon_b=n_b^2$ (assumed lossless, see Fig.~\ref{fig:film} left side), the
Green function is known analytically.
For two $s$-polarized dipoles ($z$-polarized),
\begin{equation}
G_{12}^{ss}=\frac{k_{0}^2 e^{ik_{b}r_{12}}}{4\pi r_{12}}\bigg[1+\frac{i}{k_{b}r_{12}}-\frac{1}{(k_{b}r_{12})^2}\bigg],
\end{equation}
and for two $p$-polarized dipoles ($x$-polarized),
\begin{equation}
G_{12}^{pp}=G_{12}^{ss} +\frac{k_{0}^2 e^{ik_{b}r_{12}}}{4\pi r_{12}}\bigg[-1-\frac{3i}{k_{b}r_{12}}+\frac{3}{(k_{b}r_{12})^2}\bigg],
\end{equation}
where the dipoles are separated by distance
$r_{12}$, center to center,
$k_0 = \omega/c$ and
$k_b = n_b \omega/c$. For dipoles
that are sufficiently close together (near field regime), one can approximate the above forms
with the static contribution only,
so that:
\begin{equation}
G_{12}^{ss}\vert_{\rm near}=
-\frac{1}{4\pi\epsilon_b r_{12}^3},
\end{equation}
and for two $p$-polarized dipoles,
\begin{equation}
G_{12}^{pp}\vert_{\rm near} =
\frac{2}{4\pi  \epsilon_b r_{12}^3}.
\end{equation}
These {\it near field coupling} terms, in combination
with Eq.~\eqref{eq:F}, 
identically recover the well known
F\"orster coupling rates~\cite{Frster1948}.

\vspace{0.2cm}
\noindent\textbf{Radiative decay rates and photon exchange rates for lead perovskite quantum dots}

Experiments on single 
PQDs~\cite{Rain2016}
report a single radiative lifetime of 250 ps, corresponding to a spectral linewidth of 
around 2.7~$\mu$eV. These rates are also consistent with those reported in Ref.~\cite{Becker2018}, 
for ${\rm CsPbBr}_3$ for QDs around 10-nm cubed.
The reason for these significant decay rates,
is that such QDs are in the intermediate to weak quantum confinement regime,
as the QD size is larger than the exciton Bohr radius.

For our PQDs,
the electron Bohr radius is  obtained from $s_e = 4 \pi \epsilon_{QD} \epsilon_0 \hbar^2/(e^2 m_e)  \approx 1.9~$nm (using $m_e = 0.13m_0$ and $\epsilon_{QD}=4.8$), which is clearly much smaller than the size of our QDs (10\,nm). This
large ratio is the origin of the large oscillator strength for the radiative decay~\cite{nair1997theory}, a mesoscopic enhancement effect.
This in turn enhances the radiative decay rate and  also to the dipole-dipole exchange rates.

In Ref.~\cite{Becker2018},
a variational approach was given 
with the following trial wave functions
(electron-hole picture):
\begin{equation}
\begin{aligned}
   v({\bf r}_e,{\bf r}_h)
   = C e^{-b\vert{\bf r}_e-{\bf r}_h\vert}
   \phi({{\bf r}_e}) \phi ({{\bf r}_h}),
\end{aligned}
\end{equation}
where
$C$
is determined from
\begin{equation}
    \int d{\bf r} d{\bf r}' 
    v({\bf r},{\bf r}') = 1,
    \label{eq:C}
\end{equation}
and
\begin{equation}
\begin{aligned}
   \phi({\bf r})
   = 
   \left(\frac{2}{L}\right)^{3/2}
   \cos\left(\frac{\pi x}{a}\right) \cos\left(\frac{\pi y}{a}\right)
    \cos\left(\frac{\pi z}{a}\right).\\
\end{aligned}
\end{equation}
The envelope wavefunction of the exciton $v({\bf r},{\bf r}')$ (from a product of the ground state electron and hole ground wave functions) is required to compute the oscillator strength and exciton dipole moments.

The parameter $b$ has been extracted
as a function of QD cube length $L$, and
for $L/a_e \approx 5-6$ is $b=1.75/L$ 
(see supplementary material of \cite{Becker2018}). 
Carrying out the 6D integration (Eq.~\eqref{eq:C}) with this new wave function, then the oscillator strength increases by $C$, 
which we compute to be
$C=3.79$ for $b=1.6$;
$C=4.25$ for $b=1.75$; and
$C=4.71$ for $b=2$. 
Thus we estimate that 
the excitonic oscillator strength 
is about a factor of 4-5 larger
than that for a QD in the strong confinement regime.
Using wave functions for excitons in the weak confinement regime, increases the oscillator strength even further~\cite{Becker2018}, and scale with $8L^3/\pi^2 a_B^3$, where $a_B$ is the exciton Bohr radius.

The main photoluminescence lines of PQDs consist of three linearly-polarized triplet excitons, 
which are nondegenerate, one of which is dominant.
At elevated temperatures, this will be significantly broadened due to non-radiative processes, such as electron-phonon interactions,
but these should not affect the coherent dipole-dipole interactions.

As an example, 
if we consider two coupled 
dipoles of magnitude 
$d_1 = d_2 = 0.72~{\rm e \cdot nm}$ (which give the same radiative decay rates mentioned before), as well as
$r_{12}=11~{\rm nm}$ (which includes a 1 nm gap, Fig.~\ref{fig:film} left side),
$\epsilon_b=1.5^2$,
$\hbar\omega=\hbar\omega_1=\hbar\omega_2= 2.4$~eV,
then 
\begin{equation}
\hbar\Gamma_1=\hbar\Gamma_2 = 
\frac{n_b \omega_1^3 d_1^2}{3 \pi \epsilon_0  c^3} =
2.69~\mu{\rm eV},
\end{equation}
corresponding to a lifetime of around 250~ps.
However, these estimates do not account for local field corrections, which are discussed below.
They also only consider one linearly polarized dipole, and we know there are three active dipole moments likely contributing to out main emission
resonance.

The coherent exchange term,
 with $s$-polarized dipoles ($z$-polarized),
is
\begin{equation}
\hbar\Delta_{12}^{ss}=
\frac{d^2}{4\pi \epsilon_0  \epsilon_b r_{12}^3}
\approx 93~ \hbar \Gamma_1 
=
0.25~{\rm meV},
\end{equation}
while for $p$-polarized dipoles ($x$-polarized), 
\begin{equation}
\hbar\Delta_{12}^{pp}=
-\frac{d^2}{2\pi \epsilon_0 \epsilon_b r_{12}^3}
\approx
-186~\hbar\Gamma_1 
 = -0.50~{\rm meV}.
\end{equation}

As expected, and as mentioned earlier,
these are identical in form to the well
known F\"orster coupling rate~\cite{Craig1982},
\begin{equation}
    V_F= \hbar\Delta_{12}  = \frac{d_1 d_2}{4\pi\epsilon_b \epsilon_0 r_{12}^3}
    (\delta_{ij} - \hat n_i \hat n_j),
\end{equation}
for two dipoles that are polarized along $i$
and $j$.

Without considering any effects from the metal film (or DNH topology) or local field corrections, we thus expect a maximum shift, for $p$-polarized dipoles ($x$-polarized),
of around $-186\hbar\Gamma_{1}$. This is significant, 
and estimated to be around 0.5 meV red shift 
(within a dipole approximation).

\vspace{0.5cm}
\noindent\textbf{Local field corrections}

The radiative decay rates
and photon-exchange rates
are affected by local field effects,
since the PQD cubes have a different dielectric constant than the surrounding medium.
Specifically one has~\cite{thranhardt2002relation}
\begin{equation}
    \Gamma^{\rm loc}_1 =
    D_1^2 \Gamma_1 = 
    \left(\frac{3 \epsilon_{b}}{2 \epsilon_{b}
    + \epsilon_{QD}}\right)^2\, \Gamma_1.
    \label{eq:lf}
\end{equation}
Values for $\epsilon_{QD}$ on PQDs vary in the literature
from around
6-7~\cite{shcherbakovWu2021temperature}
in bulk (and thin films) to 4.8 for QDs~\cite{Becker2018}.
So we will consider two values,
 $\epsilon_{QD}=6.25$
 and  $\epsilon_{QD}=4.8$.
 With $\epsilon_{QD}=6.25$, then 
 we expect a local field reduction factor
of
\begin{equation}
\Gamma_1^{\rm loc} \approx 0.4 \Gamma_1.
\end{equation}
This is only an estimate
as it neglects finite size effects,
but this is expected to be reasonable for very
small PQDs, essentially only
including a depolarization factor.
If we use $\epsilon_{QD}=4.8$, then the local field reduction is slightly less, with
$D_1^2 = 0.53$. 

Since we have estimated
$d=d_1=d_2$ from the experimental radiate decay rates, clearly these
are reduced from the actual ones, but it does not  matter if we regard
$d$ as an {\it effective dipole moment} including
local field corrections, or we can simply increase
this value in the presence or local field effects
(which would reduce the local all rates).
The reason this does not matter,
is that for identical QD cubes, the same reduction factor applies
also to the dipole-dipole coupling rates
\cite{PhysRevA.75.042109}. Thus,
\begin{equation}
 \frac{{\rm Re} G_{12}\vert_{\rm loc}}{{\rm Im} G_b\vert_{\rm loc}} \approx
    \frac{{\rm Re} G_{12}}{{\rm Im} G_b},
\end{equation}
and so we can once again scale our frequency shift in terms on the experimental radiative decay rate.
We will show the explicitly below,
with full numerical simulations.

\vspace{0.2cm}
\noindent\textbf{Exact photon Green function solutions
in the presence of a gold film}

To more accurately assess  the effects
of the gold film, and to include
the possibility of plasmon polariton effects and image dipole effects,
we now calculate the exact Green functions
with a metal film, specifically a 70-nm gold film (Fig.~\ref{fig:film} right side).
Clearly this is only an estimate for our DNH structure,
which may also have some background resonance effects 
from the nano-hole geometry. Nevertheless, it is useful to assess the expected role of the gold film.

We use a well established multi-layered Green function 
approach~\cite{PhysRevE.62.5797}. For gold, we 
adopt a simple Drude model for the complex dielectric constant
\begin{equation}
\epsilon_{\rm film} =    \epsilon_{\rm gold}
= 1 -\frac{\omega_p^2}{\omega(\omega+i \gamma)},
\end{equation}
with $\gamma=1.41 \times 10^{14}~{\rm rads/s}$
and $\omega_p=1.26 \times 10^{16}~{\rm rads/s}$.
However, we also consider a more realistic model~\cite{johnson_optical_1972}.

\begin{figure}[hbpt]
  \centering
  \includegraphics[width=0.75\columnwidth]{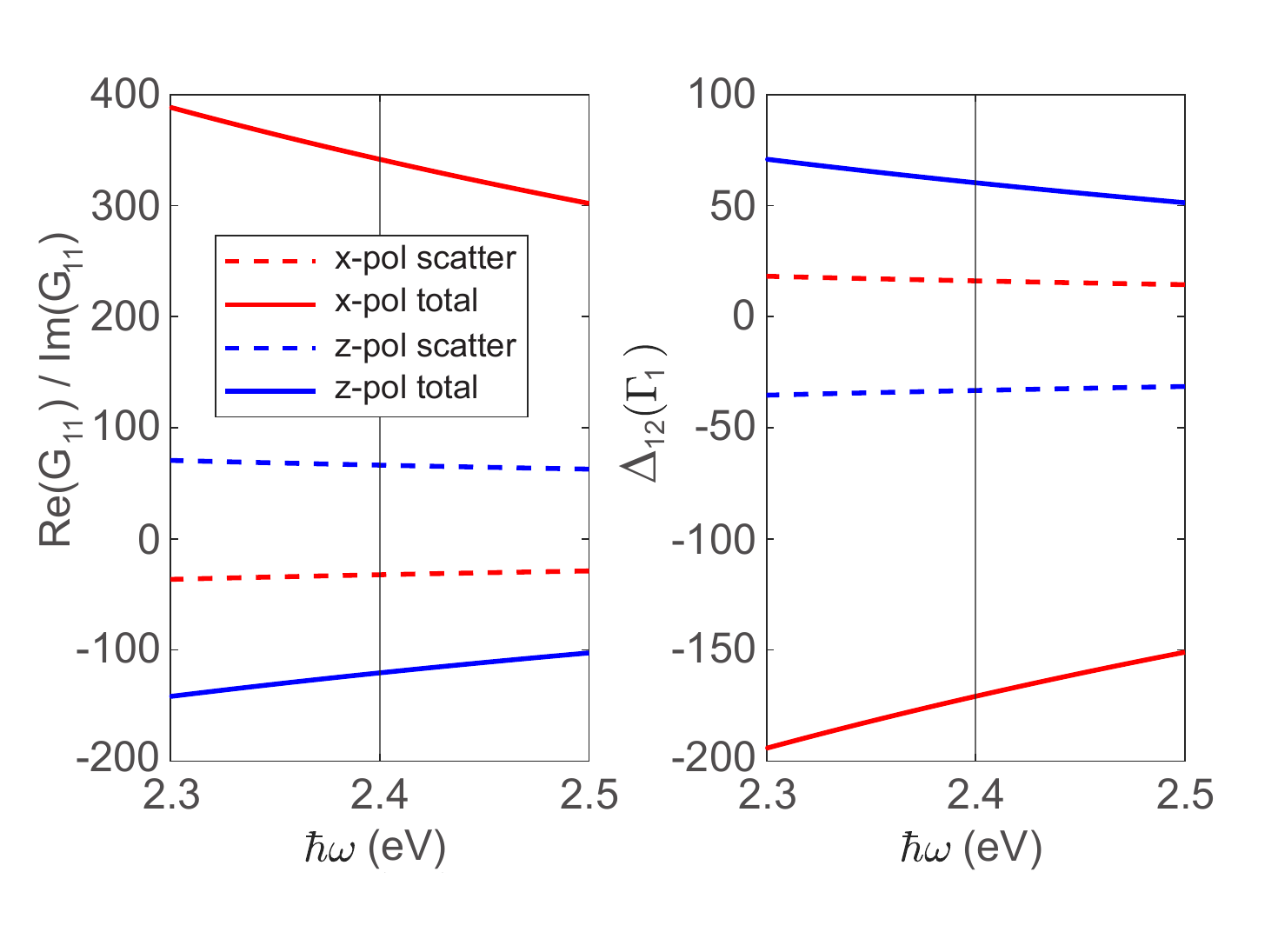}
  \caption{Green function calculation of the propagators (in Purcell factor units),
  and corresponding frequency shifts from dipole-dipole coupling. We use
  two dipoles, 6\,nm above a metal film, separated in $x$
  by 11\,nm (Fig.~\ref{fig:film} right side). 
  Note the scattered components from the metal film contribute negatively, and thus reduce the usual F\"orster coupling. The thin vertical line corresponds to a frequency close to 
  our experiments.
  }\label{fig:dd1}
\vspace{0.1cm}
  \includegraphics[width=0.75\columnwidth]{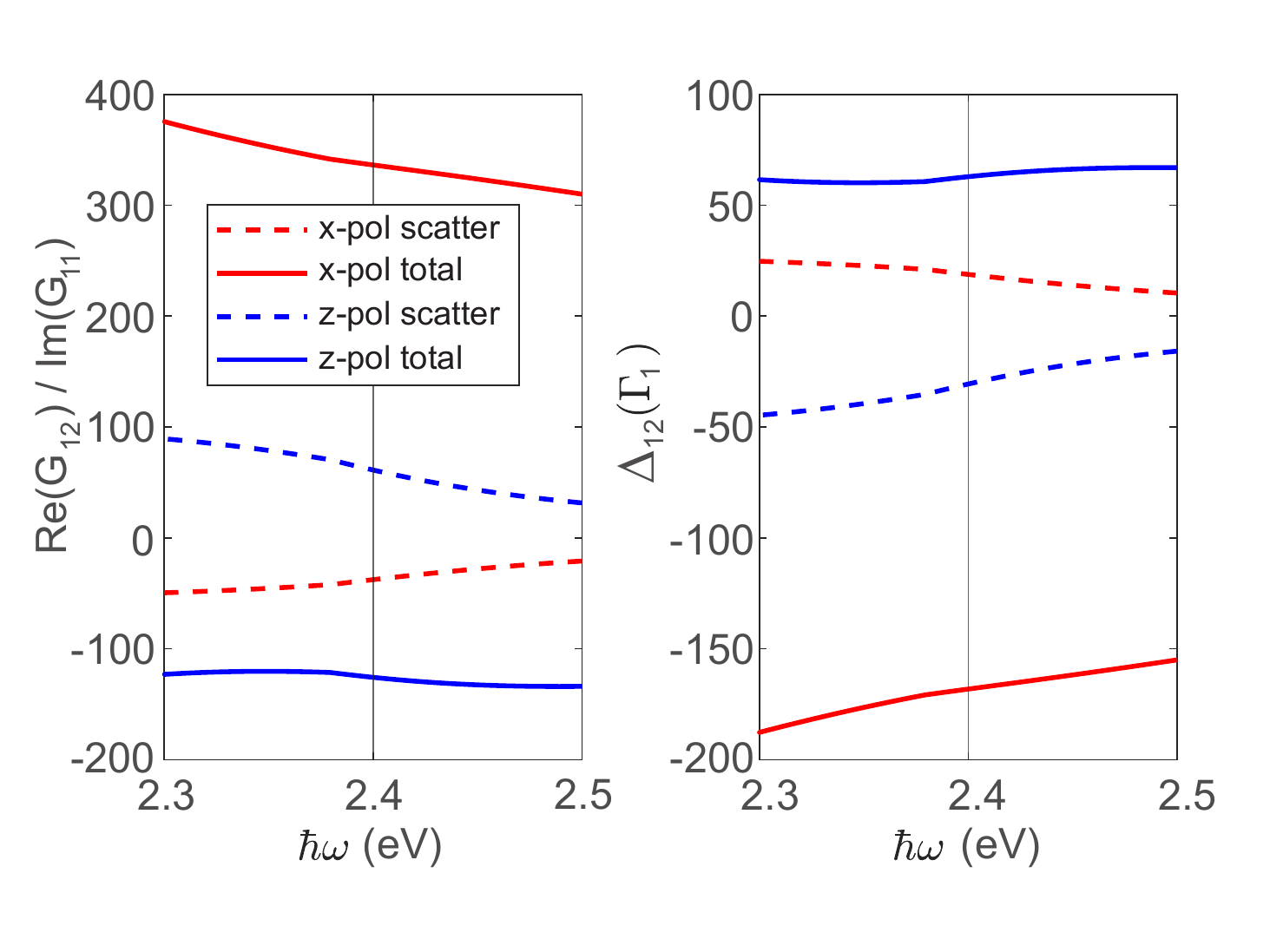}
  \caption{Same as in Fig.~\ref{fig:dd1}, but using Johnson and Christy data for gold~\cite{johnson_optical_1972}. As can be seen, there are little qualitative differences in the frequency of interest.
  }\label{fig:dd2}
\end{figure}

Fig. \ref{fig:dd1} shows the computation of the propagator for coupling two dipoles (Drude model),
that are 6 nm from the metal (1-nm height/gap above the interface, Fig.~\ref{fig:film} right side), which act to reduce the homogeneous contributions given above. For example, the
real part of the $s$-polarized (z-polarized) propagator (${\rm Re}\{G_{12}^{\rm total}\}$) changes from 
-182 (using the analytical Green functions, with a homogeneous medium, in Purcell factor units) to -115 (with a gold film included in the calculations, solid blue curve at $\hbar\omega=2.4~$eV in left side of Fig. \ref{fig:dd1}).
We have also computed the same function in COMSOL and obtained -114 (not shown), a remarkably good agreement. Fig.~\ref{fig:dd2} shows identical simulations,
but with  a more realistic 
material model for gold~\cite{johnson_optical_1972}. Clearly, we obtain exactly the same trends, so the Drude model seems reasonable for these simulations.

\vspace{0.2cm}
\noindent\textbf{Full electromagnetic simulations using COMSOL}

To corroborate the above analytical studies and arguments,
we also compute the photon Green functions numerically,
for a general medium including the QD cubes and the metal film. 
This can also give insight into any extra scattering effects that may occur, such as scattering from the 
coupled QD cubes, though we expect such effects to play a minor role. 
We use $\epsilon_b=2.25$ and
$\epsilon_{QD}=n_{\rm cube}^2=6.25$, but the results will scale
with different $\epsilon_{QD}$, since we divide the
final photon exchange terms by 
calculated radiative rates that also include local field corrections.
We have also checked this explicitly,
and the relative differences with  $\epsilon_{QD}=4.8$ compared to 6.25
are only a few percent.

\begin{figure*}[hbpt]
  \centering
  \includegraphics[width=0.84\columnwidth]{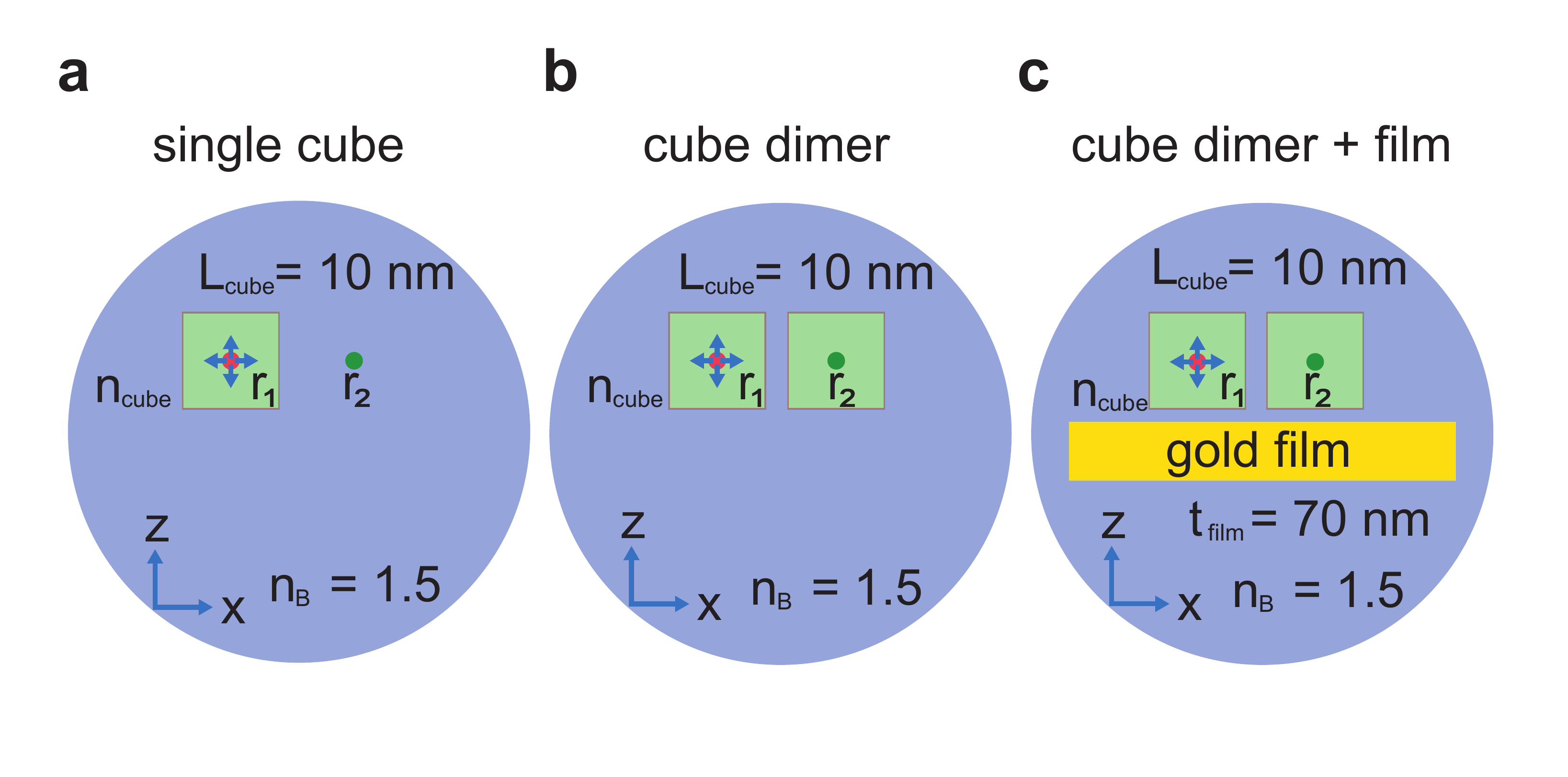}
  \caption{Schematic diagram for single cube, cube dimer (side by side, here the surface to surface gap is $1$~nm) and cube dimer plus the gold film (cube dimer is placed $1$~nm above the gold film), used in the numerical COMSOL simulations. The source dipole is placed in the center $\mathbf{r}_{1}$ of the left cube, where we consider $x$- or $z$-polarized.
  }\label{fig:cubedimer_sche_new}
\end{figure*}

To compute the numerical Green function for a general photonic medium, we use COMSOL, including the PQD cubes (modified background response) and the gold film.
With a dipole source excitation, we could easily get $\mathbf{E^{\rm total}}(\mathbf{r})$ and $\mathbf{E^{\rm B}}(\mathbf{r})$, which are the electric fields with and without the scatterers in the medium. The scattered field is then obtained from $\mathbf{E^{\rm scatt}}(\mathbf{r})=\mathbf{E^{\rm total}}(\mathbf{r})-\mathbf{E^{\rm B}}(\mathbf{r})$.
To obtain the numerical scattered Green function, we exploit the following expression for the scattered field from a dipole source at $\mathbf{r}_{0}$,
\begin{equation}\label{eq: EfromG}
 \mathbf{E^{\rm scatt}(r,\omega)}=\mathbf{G}^{\rm scatt}(\mathbf{r},\mathbf{r}_{0},\omega)\cdot\frac{\mathbf{d}}{\epsilon_{0}},  
\end{equation}
where the scattered field $\mathbf{E^{\rm scatt}}$ has units of $\rm V/m$, the Green function $\mathbf{G}^{\rm scatt}$ has units of ${\rm m}^{-3}$,  and the dipole moment $\mathbf{d}$ 
has units of $\rm C\cdot m$.
We can also obtain the total Green function in the same way,
\begin{equation}\label{eq: EfromG2}
 \mathbf{E^{\rm total}(r,\omega)}=\mathbf{G}^{\rm total}(\mathbf{r},\mathbf{r}_{0},\omega)\cdot\frac{\mathbf{d}}{\epsilon_{0}},  
\end{equation}
though one should avoid computing 
${\rm Re}\{{\bf G}^{\rm total}({\bf r}_0,{\bf r}_0)\}$, as it is divergent;
however, ${\bf G}^{\rm scatt}({\bf r}_0,{\bf r}_0)$ (real and imaginary parts)  and all other Green function quantities ($\mathbf{r}\neq\mathbf{r}_{0}$) are well defined.

\begin{figure}[hbpt]
  \centering
  \includegraphics[width=0.75\columnwidth]{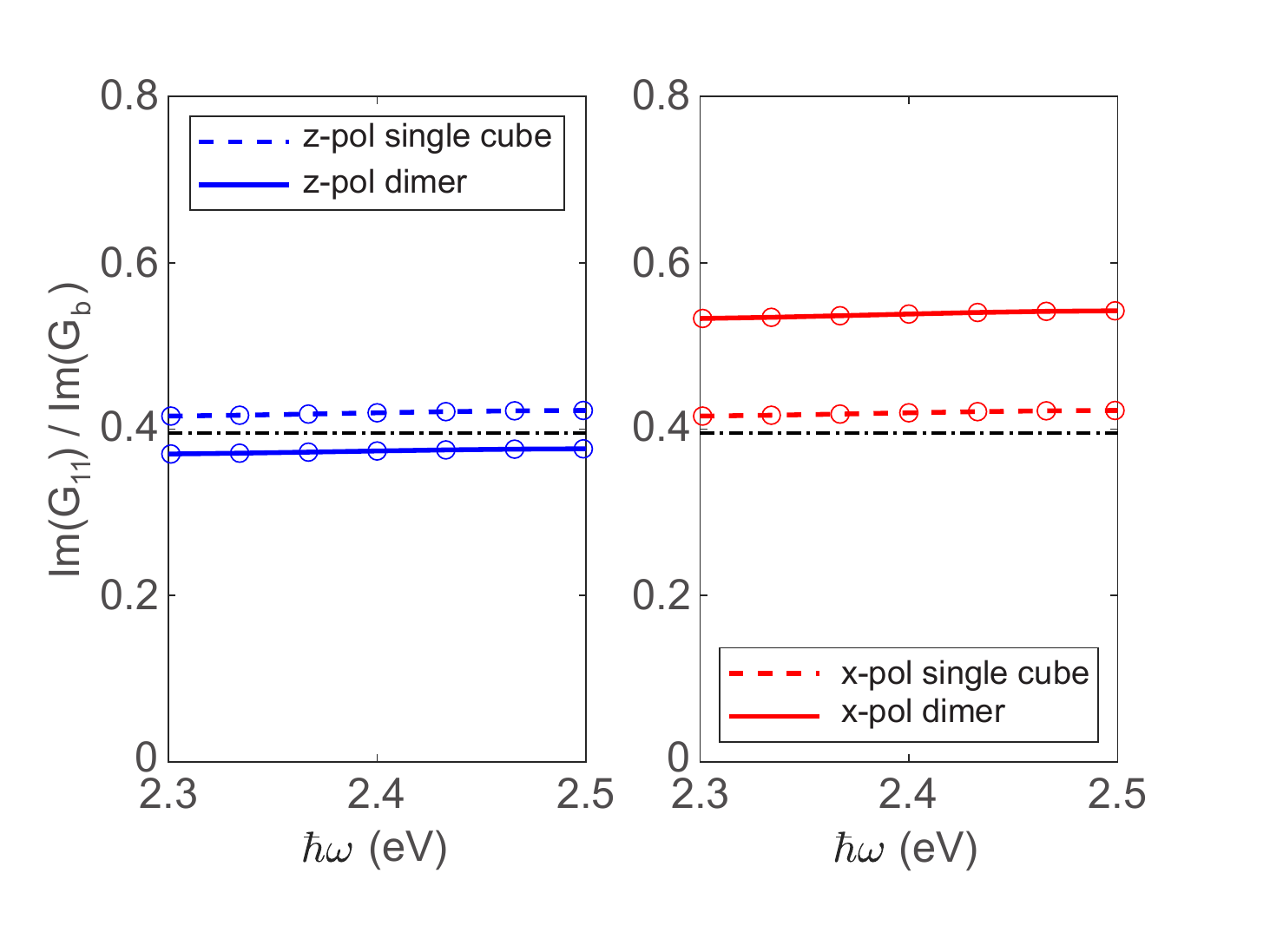}
  \caption{Projected LDOS in Purcell factor units, showing the effect of local field corrections with a single cube and a cube dimer (Fig.~\ref{fig:cubedimer_sche_new} (a,b)).
  The expected analytical answer is also shown for a single cube (Eq.~\eqref{eq:lf}, thin chain line).
  Note the lines and symbols are the same data, but more accurately show the numerically computed data points (symbols).
  }\label{fig:LDOS}
\end{figure}
\begin{figure}[hbpt]
  \centering
  \includegraphics[width=0.75\columnwidth]{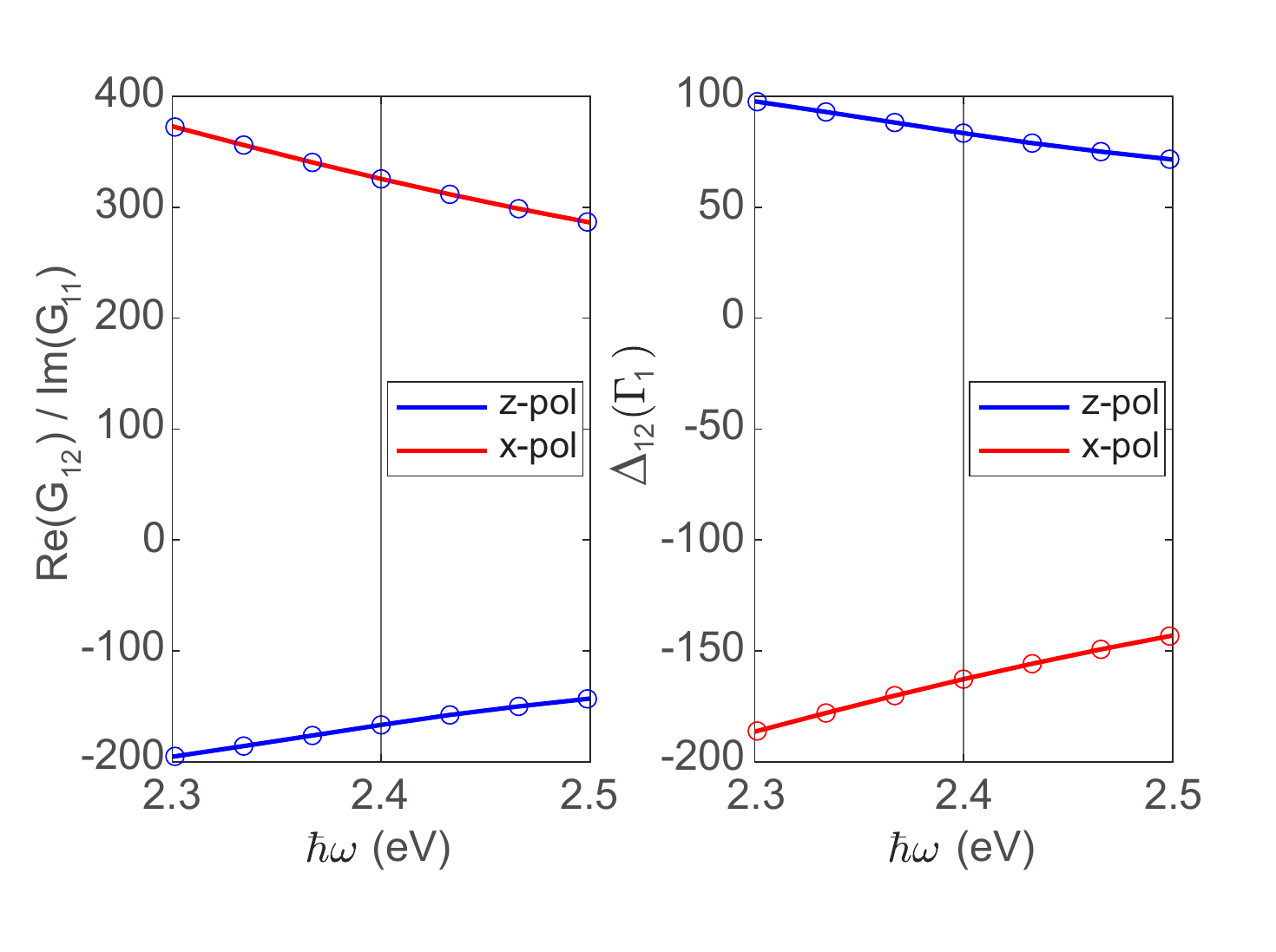}
  \caption{Propagators and dipole-dipole shifts computed from COMSOL, which includes the full scattering geometry (Fig.~\ref{fig:cubedimer_sche_new} (c)). All terms include local field corrections, so are normalized by the result including the single QD cube. The decay rate, $\Gamma_1$, is thus the one also including local field corrections for a single QD cube. The vertical line is close to the experimental frequency regime.
  }\label{fig:Prop}
\vspace{0.5cm}
  \includegraphics[width=0.73\columnwidth]{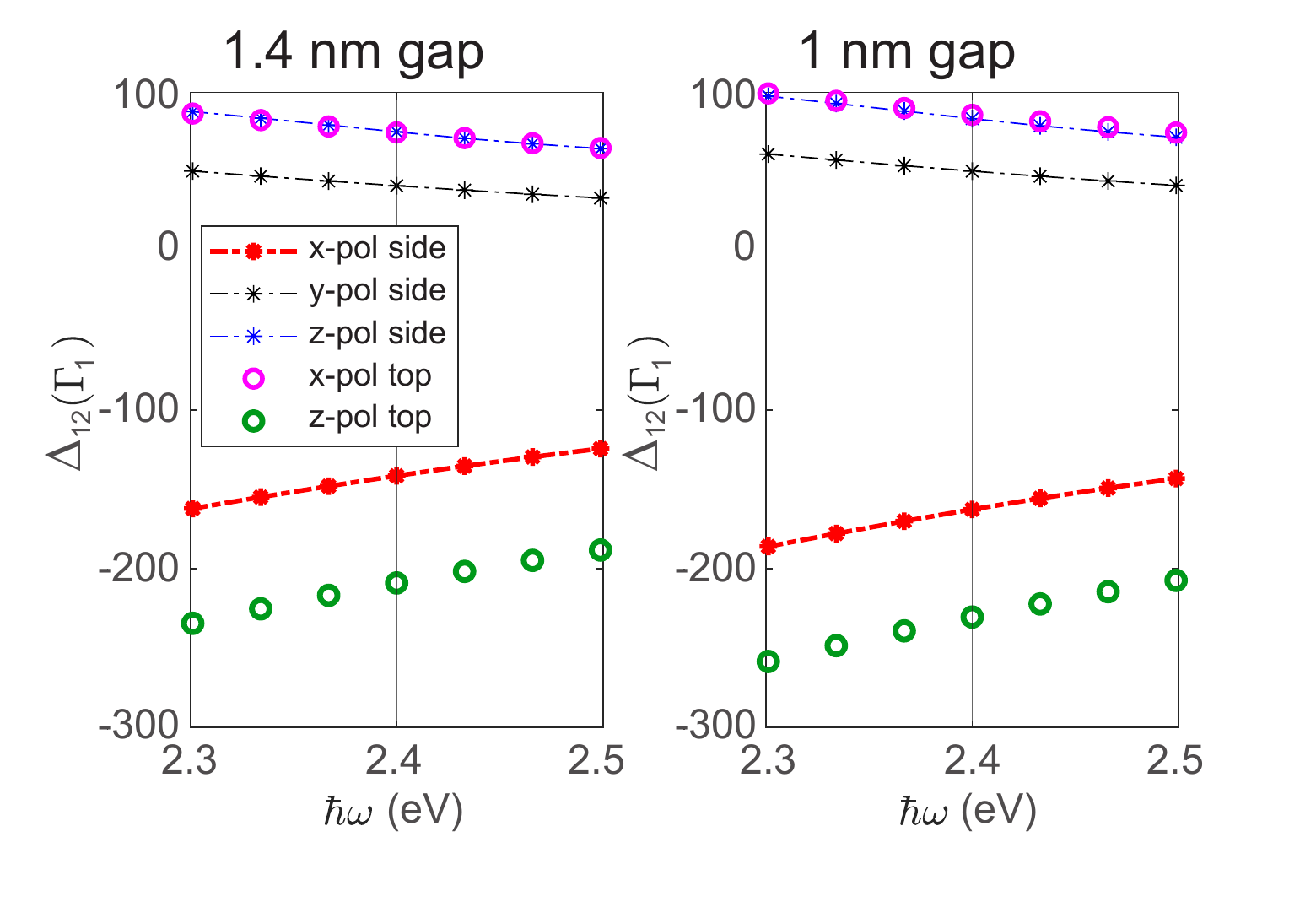}
  \caption{Dipole-dipole frequency shifts computed in COMSOL, showing the difference between $1$~nm (right side, same as Fig.3 c right figure in main text) and $1.4$~nm gaps (left side). The larger gap causes a very small reduction, as expected. With $1.4$~nm gap, the maximum photon exchange rate is around $-209\hbar\Gamma_1$ for $z$-polarized dipoles with top coupling, which we estimate to be around $-0.56$~meV. } \label{fig:d14}
\end{figure}

Within the simulation, the source dipole is placed at $\mathbf{r}_{1}$ (Fig.~\ref{fig:cubedimer_sche_new}) and assumed polarized along some $i$ direction (such as the $x,y,z$). 
Thus we can write the dipole moment  as $\mathbf{d}=d\mathbf{n}_{i}$, where $d$ is the value of the dipole moment, and $\mathbf{n}_{i}$ is a uniot vector. Note this dipole moment is not related to $d_1$ and $d_2$, but is simply
a linear dipole in Maxwell's equations to obtain the numerical Green functions.

First we explore local field corrections, with one PQD and 
two PQDs (Fig.~\ref{fig:cubedimer_sche_new} (a,b)), since additional scattering can be captured that may show up on top of the usual formula shown earlier, namely
 Eq.~\eqref{eq:lf}. Fig. \ref{fig:LDOS}
shows the COMSOL simulations from the PQD cubes. The single PQD results are shown
to match the analytical expressions very well, and the two PQD case shows a departure in the case of $x$-polarized dipoles ($p$-polarized). Thus, additional scattering is shown to enhance the projected LDOS (local density of states), or
for an $x$-polarized dipoles; however, its impact is relatively small.

Next, to connect to the
photon exchange coupling rates (which will include the usual F\"orster contribution),
we compute the
numerically exact propagator (within numerical uncertainties), and corresponding scattering rates (for cube dimer placed side by side above a gold film, Fig.~\ref{fig:cubedimer_sche_new} (c)),
as shown in Fig.~\ref{fig:Prop}. Importantly, these are normalized by the single PQD decay rate, and all simulations include the PQDs; thus local field effects are also included in the normalization, and all scattering effects are included for the calculation of the propagators.

Overall, the trends are seen to be in good agreement with the analytical results shown earlier in Fig.~\ref{fig:dd1}. 
Note that if we remove the PQD cubes from the COMSOL calculations and include only the metal film (Fig.~\ref{fig:film} right side), then the agreement with the
analytical solution is less than 1\%. As mentioned above, the real part of the total Green functions ${\rm Re}(G_{12})$ for an $z$-polarized dipole
is $-115$ (in Purcell units, solid blue curve at $\hbar\omega=2.4~$eV in left side of Fig. \ref{fig:dd1}) from the analytical Green function results and $-114$ from COMSOL simulations (not shown).
Even with small
quantitative differences 
the overall conclusion is the same.
When with full structure, i.e., PQDs separated by 11-nm (placed side by side), and 1-nm above a gold film (Fig.~\ref{fig:cubedimer_sche_new} (c)), the maximum frequency shift expected
is around $-163\Gamma_1$, where $\Gamma_1$ includes local field corrections. Assuming
$\hbar\Gamma_1 \approx 2.69~\mu$eV, then we predict 
a spectral shift of $-0.44~$meV, for
$p$-polarized ($x$-polarized) dipoles.

Additional calculations for vertically coupled QD cubes on gold film (one cube is placed on the top of the other one) are shown in the main text.

As mentioned in the main text, after initial calculations at a $1$~nm gap size, we obtained a more accurate measurement of the gap size from high-resolution TEM as being $1.4$~nm. This is expected theoretically to reduce the interaction by around $10\%$, which we confirmed by simulation of an typical structure (see Fig.~\ref{fig:d14}).

\vspace{0.5cm}
\noindent\textbf{Other potential sources and mechanisms for coupled-PQD frequency shifts}

Next, we briefly discuss some other potential effects that could cause large frequency shifts for coupled
PQDs.

For quantum emitters that are very close together, namely, where electronic tunneling is possible, 
 excitons can be transferred,
 which is typically termed
Dexter energy transfer~\cite{DuBose2021}.
This process is usually negligible for distances greater than 1~nm.
Dexter is a non-radiative process with electron exchange. Although similar to F\"orster energy transfer, if differs greatly in length scale and the underlying mechanism.
Dexter transfer can be singlet-singlet or
triplet-triplet,
and the three fundamental excitations
in lead PQDs are triplets~\cite{Becker2018}.
Dexter energy transfer is a process where the donor and the acceptor exchange their electron. Thus, the exchanged electrons should occupy the orbital of the other pair.
The Dexter transfer rate has the form
$ V_{D} \propto J \exp(-2r_{12}/L_v)$,
where $J$ is the normalized spectral overlap,
and $L_v$ is the the sum of van der Waals radius.
Dexter transitions couple bright excitons to bright excitons and are spin preserving. Their strength can also be computed from the Coulomb matrix element.
However, since Dexter coupling requires overlap with the electronic wave functions, we believe it is highly unlikely for our coupled QDs.

There are also potential monopole-monopole interactions,
which do not depend on any spatial overlap
of the electronic wave functions. These are usually neglected
in the derivation of F\"orster coupling between
quantum emitters. Actually, monopole-monopole
is also the origin of Dexter energy transfer, but in that case, the contribution also requires oribital overlap.
In the linear 
excitation domain (e.g., neglecting biexciton effects),
the excitonic monopole-monopole term~\cite{Richter2006,PhysRevB.91.155313} 
 merely renormalizes the eigenergies, unlike F\"orster and Dexter
terms which represent exchange of photons or electronic excitations. 
Furthermore, 
for a symmetric wavefunctions, monopole-monopole shifts should be zero for linear excitation.

There could also be effects beyond a dipole approximation, which can be captures by doing a spatial integration with respect to the inter-PQD exciton wave functions and the photon Green function propagator \cite{Carlson2020}. We have carried out such a calculation using 6D Monte Carlo integration and found the dipole approximation to be excellent,  within 5\% for QD cubes that have a gap separation of 1~nm. This calculation used the ground
state exciton wave function, and we used a similar integration techniques to compute $C$, which enhances the oscillator strength.

Finally, we mention tandem tunneling~\cite{Reich2016} that can occur through the intermediate state in which the electron and hole are in different PQDs. This process 
has been shown to have exciton hopping rates that are larger than the Dexter rate F\"orster 
for certain PQDs. However, such PQDs are
clearly fused together (unlike our optically trapped coupled PQDs), so we also rule 
out the effect of fused tunneling.

\vspace{0.5cm}
\noindent\textbf{Optical binding}

By using the Rayleigh theory in Eq.~\ref{rayleigh} for the scattering cross section~\cite{Pierrehumbert2014}, the polarizability constant ($\alpha_p$) of the scatterer can be calculated from the scattering cross section:

\begin{equation}
    x_{scat} = \frac{8\pi}{3}\frac{(2\pi)^4}{\lambda^4}\alpha_p^2
    \label{rayleigh}
\end{equation}
 
The trapping potential energy in the dipole limit can be given by\cite{gordon2022future}:
\begin{equation}
    U_{trap} = -\frac{1}{2}\vec{p}\cdot\vec{E}
    \label{trap_potential}
\end{equation}

\noindent where $\vec{p}$ is the dipole moment and $\vec{E}$ is the electric field. The optical potential energy should be greater than the thermal energy to keep trapping the particle.

\begin{figure}[htbp]
  \centering
  \includegraphics[width=120mm]{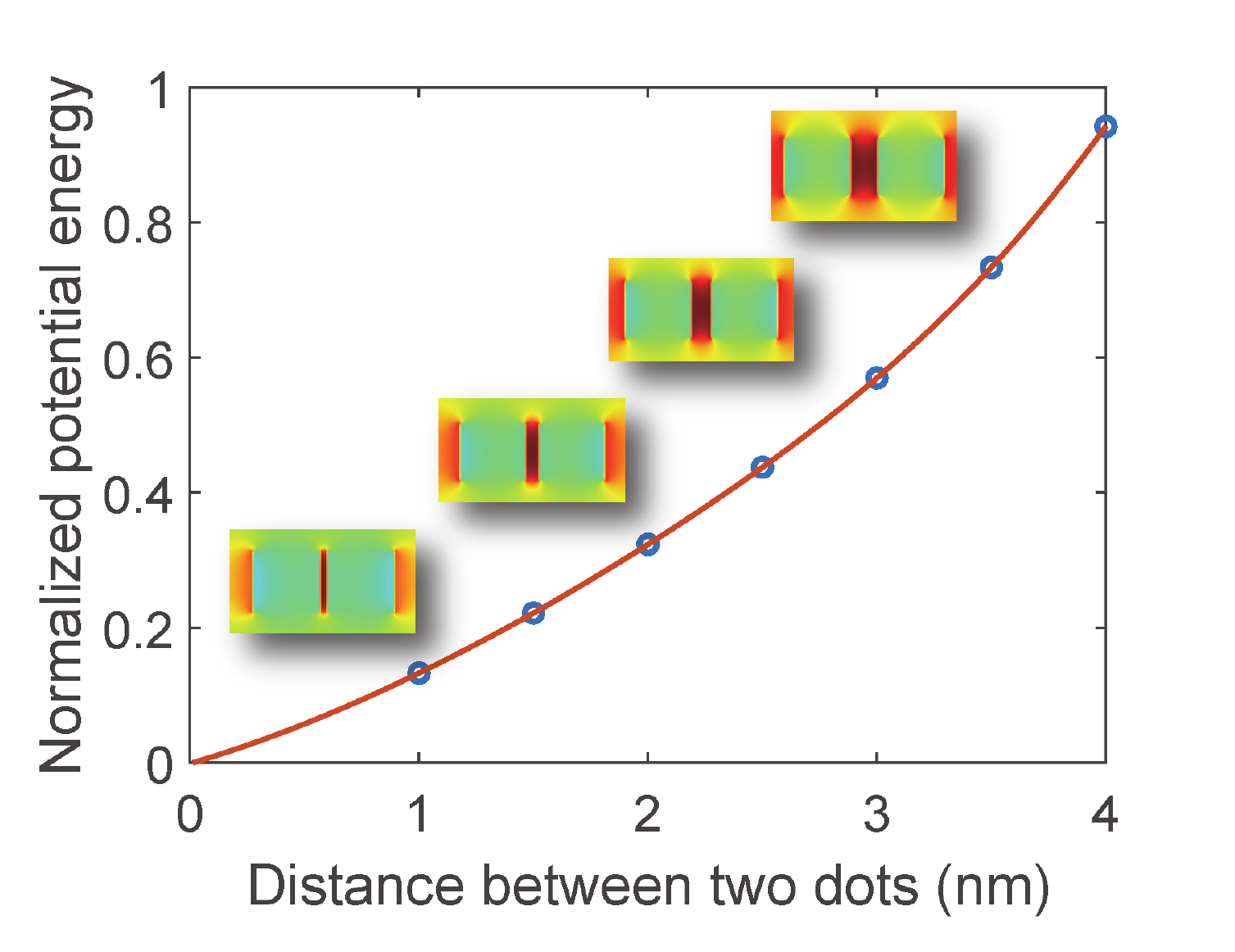}
  \caption{FDTD simulation results of the trapping potential energy against the distance between two dots. Inset figures show the visualization of the electric field intensity of the double dots with various distance.
  }\label{FDTD}
\end{figure}

Here, we use the finite-difference time domain simulation software (Lumerical FDTD ver. 2020 R2.3). The simulations calculated the trapping potential energy of the single and double PQDs (dielectric cubes, $\epsilon=4.8$). Fig. ~\ref{FDTD} shows the potential energy as a function of the distance between two PQDs. The electric fields with 1 nm - 4 nm are shown in inset figures. This shows that the dots are attracted to one another through the optical field in the dipole limit.

\clearpage
\vspace{1cm}

\bibliography{achemso-demo}

\end{document}